\documentclass[%
 reprint,
 amsmath,amssymb,
 prx,
]{revtex4-2}
\usepackage[utf8]{inputenc}

\usepackage{graphicx}

\usepackage{xcolor}
\usepackage{amsmath}
\usepackage{physics}

\usepackage[percent]{overpic}

\usepackage[hidelinks]{hyperref}
\usepackage[caption=false]{subfig}

\newcommand{\olo}{\mathrm{LO}}

\newcommand{\uvmshz}{\,\mathrm{\mu V\,m^{-1}\,Hz^{-1/2}}}

\newcommand{\EA}{E_\mathrm{RF}}
\newcommand{\bEA}{\boldsymbol{E}_\mathrm{RF}}
\newcommand{\wsig}{\omega_\mathrm{RF}}
\newcommand{\ER}{E_E}
\newcommand{\Sro}{S_{11}}
\newcommand{\Srt}{S_{22}}
\newcommand{\Sko}{S_{12}}
\newcommand{\Skt}{S_{21}}
\newcommand{\Ssig}{p_{\mathrm{sig}}}
\newcommand{\Sa}{p_{A}}
\newcommand{\Sl}{p_{L}}
\newcommand{\Sp}{p_{p}}
\newcommand{\satt}{\xi}
\newcommand{\fsig}{f_s}
\newcommand{\diplus}{d_{i}^{+}}
\newcommand{\diminus}{d_{i}^{-}}
\newcommand{\diplusminus}{d_{i}^{\pm}}

\newcommand{\nth}{n_\mathrm{th}( \fsig ,T)}

\begin{document}

\title{Comparison of Noise Temperature of\\ Rydberg-Atom and Electronic Microwave Receivers }	
\author{Gabriel Santamar\'ia Botello$^1$}
\author{Shane Verploegh$^2$}
\author{Eric Bottomley$^2$}
\author{Zoya Popovi\'c$^1$}
\affiliation{$^1$ University of Colorado, Boulder, U.S.A.}
\affiliation{$^2$ ColdQuanta, Inc., Boulder, U.S.A.}
\date{\today}
	
\begin{abstract}

Microwave receivers using electromagnetically-induced transparency (EIT) in Rydberg atoms have recently 
demonstrated 
improved sensitivities. 
It is not 
evident how their state-of-the-art electric field sensitivities compare to those achieved using standard electronic receivers consisting of low-noise amplifiers (LNAs) and mixers. 
In this paper, we show that conventional room-temperature electronic receivers greatly outperform the best demonstrated sensitivities of room-temperature Rydberg electrometers in standard free-space coupled configurations. However, Rydberg-atom receivers can surpass the sensitivity of conventional receivers if 
resonant or confining microwave structures are designed to enhance the electric fields sensed by the atoms. 
For a given microwave resonator, 
the external (coupling) quality factor 
must be carefully chosen to minimize their thermal and quantum noise contributions. Closed-form expressions for these optimal design points are found, and compared in terms of noise temperature with conventional LNAs reported in the literature from $600\,\mathrm{MHz}\sim 330\,\mathrm{GHz}$.


\end{abstract}
	
	\maketitle

\section{Introduction}
	
Low-noise high-sensitivity microwave receivers are a part of every communication, radar and radiometer system. A room-temperature or cryogenic low-noise amplifier (LNA) is the first receiver element that immediately follows the antenna to reduce the amount of added noise and not degrade sensitivity. Low-noise amplifiers have a fundamental limitation with transistor scaling to higher frequencies in the millimeter-wave range \cite{Marion-NoiseLimit}. Exploration of the effect of blackbody radiation on Rydberg states in 1980 led to the proposed use of Rydberg atoms as {millimeter-wave 
power detectors} with an inferred noise-equivalent power (NEP) of $10^{-17}$\,$\mathrm{W}\,  \mathrm{Hz}^{-1/2}$ \cite{HWalter-1980}. The recent increased interest in these sensors is prompted by the calculated atomic shot-noise limit in the $\mathrm{n}\mathrm{V}\,\mathrm{m}^{-1}\,\mathrm{Hz}^{-1/2}$ 
range \cite{gordon2014millimeter,Fan_2015} and predictions of exceeding performance of standard receivers. Here we present an in-depth study of the noise limitations of field-enhanced Rydberg atom microwave and millimeter-wave receivers, concluding with a fair comparison with existing LNAs.


\begin{figure}[t!]
	\includegraphics[width=0.43\textwidth]{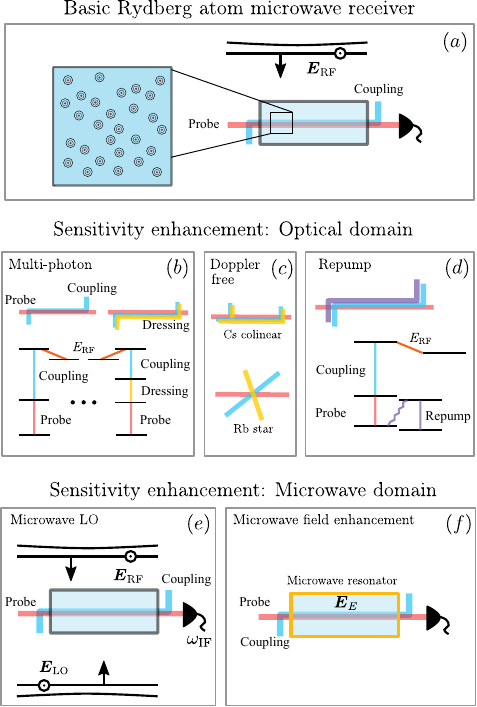}
		\caption{{Incoherent Rydberg-atom RF electric field sensor (a). Optical methods to improve sensitivity, including using multiple lasers (b), optical beam geometries (c) and an additional repump laser (d). Microwave methods for improving sensitivity include adding a microwave local oscillator (LO) as shown in (e) and microwave field enhancement (f).} }
		\label{fig:simple}
\end{figure}


The most basic room-temperature Rydberg-atom electrometer retrieves the absolute magnitude of the microwave electric field through a spectroscopic measurement of the Autler-Townes splitting of an electromagnetically-induced transparency (EIT) signature \cite{sedlacek2012microwave}. This type of electrometer, illustrated in Fig. \ref{fig:simple}(a), can be operated as an incoherent radio-frequency (RF) receiver in the small-signal regime. 
Sensing in this regime is limited by the photon shot noise of the probe laser and coupling laser intensity. Both optical homodyning and frequency modulation of the probe beam are shown to reduce noise and bring it closer to the probe shot-noise limit. 
With an all-optical demonstration in a two-laser (2-photon) system illustrated in the left of Fig.\,\ref{fig:simple}b, a sensitivity of 300 $\mu \mathrm{V} \mathrm{m}^{-1}  \mathrm{Hz}^{-1/2}$ is demonstrated in \cite{kumar2017atom}, using either homodyne detection or frequency-modulation spectroscopy. 
To further improve system sensitivity, recent 
proposals include three-photon excitation for eliminating Doppler-broadening in a co-linear optical and `star' optical configuration for Cesium and Rubidium receivers 
(Fig.\,\ref{fig:simple}c), with the colinear optical configuration in Cs showing improved optical spectral resolution 
\cite{ryabtsev2011doppler,shaffer2022polarization}. 
An additional laser that re-populates the intended Rydberg states by forcing atoms in other states to decay back to the ground state also increases the number of interrogated atoms  \cite{prajapatirepump}, further improving sensitivity (Fig.\,\ref{fig:simple}d). 


The addition of a microwave local oscillator (LO) in a Rydberg atom sensor (Fig.\,\ref{fig:simple}e) allows both amplitude and phase detection and biases the system to an high responsivity point. This was first shown in \cite{Simons-heterodyne2019}, with a record sensitivity demonstrated shortly after in \cite{jing2020atomic}. The same method is used in a recent experiment at 10.68\,GHz with an electric field detection state-of-the-art sensitivity of $1.25\,\uvmshz$ \cite{cai2022}. This coherent receiver is now directly analogous to a heterodyne microwave electronic receiver. A complementary sensitivity improvement approach is the addition of RF structures that confine or enhance the incident RF field (Fig.\,\ref{fig:simple}f). Demonstrated examples include locating the atomic sensing region close to the surface of a co-planar waveguide inside a vacuum chamber \cite{MeyerPhysRevApplied.15.014053}, embedding a parallel-plate resonator inside the vapor cell providing a field enhancement factor of $16$ at $4.35\,\mathrm{GHz}$ \cite{andersoncavitydoi:10.1063/1.5038550}, and placing the vapor cell inside a split-ring resonator to achieve a resonant field enhancement factor of 100 at 1.3\,GHz \cite{Hollowaysplitringdoi:10.1063/5.0088532}. 

A sensitivity comparison between coherent atomic and mixer-based electronic receivers is not straightforward. The latter is characterized by noise figure (NF) or noise-equivalent temperature ($T_\mathrm{noise}$ or NET), whereas the atomic sensor sensitivity is more appropriately quoted as a noise-equivalent field (NEF) in units of $\mathrm{V}\, \mathrm{m}^{-1} \,\mathrm{Hz}^{-1/2}$. One could square the NEF to convert to NEP, but this requires a definition of a specific antenna aperture, as discussed 
in \cite{Meyer_2020,fancher}.  The goals of this paper are to: (1) define antenna aperture in two ways that lead to fair comparisons between Rydberg-atom and electronic receivers; (2) define a noise temperature of a port-coupled atomic system that is independent of the specific antenna aperture and therefore comparable to LNA-mixer receiver noise temperature; and (3) formulate a tradeoff for the port coupling that minimizes noise and maximizes sensitivity. 


The paper is organized as follows. The next section presents a brief overview of Rydberg atom sensing for context and notation, 
introduces free-space and port-coupled atomic receivers and derives quantities that allow a fair sensitivity comparison with conventional receivers. 
Thermal and quantum microwave noise contributions are quantified in sections \ref{sec:free} and \ref{sec:mode}, in free-space (Fig. \ref{fig:simple}(e)) and port-coupled (Fig. \ref{fig:simple}(f)) 
configurations respectively. Port-coupled high-Q resonators are analyzed using a single-mode harmonic oscillator model, whereas a 
more general arbitrary waveguide model is derived for multi-mode or non-resonant microwave structures, describing most practical scenarios for field enhanced atomic receivers. In Section \ref{sec:results}, an example is given for an X-band waveguide resonator, followed by a numerical validation of the waveguide model using the harmonic oscillator model as a reference. 
Then, a comparison between the sensitivity of optimal field-enhanced atomic receivers and LNAs ranging from $600\,\mathrm{MHz}$ to $ 330\,\mathrm{GHz}$ is performed. 
Finally, a discussion on the usage of the theory to various microwave topologies followed by conclusions are given in Section \ref{sec:conclu}.

\section{Generalized Rydberg-Atom Microwave Sensor}
	
\begin{figure}[t]
		\centering
		\includegraphics[width=0.45\textwidth]{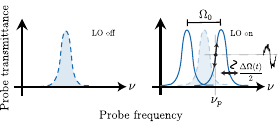}
		\caption{EIT signature after exciting atoms to Rydberg states (left) and the AT splitting induced by the presence of an on-resonance microwave LO electric field (right).}
		 \label{fig:EIT}
	\end{figure}

A simplified setup of an atomic RF sensor is shown in Fig. \ref{fig:simple}(e). Coupling  
and probe laser beams overlap inside a vapor cell filled with alkali atoms. The illuminated atoms are coupled to a Rydberg state, causing electromagnetically-induced transparency (EIT) for the probe laser at the resonance frequency, Fig. \ref{fig:EIT} (left). The presence of a microwave electric field at a frequency $\omega_\olo$, that matches the transition frequency of a neighboring Rydberg state, causes Autler-Townes (AT) frequency splitting ($\Omega_0$) of the transmitted EIT signal, proportional to the magnitude of the microwave electric field and the dipole moment of the transition, Fig. \ref{fig:EIT} (right). An optimal value $\Omega_0$ is accomplished by choosing the amplitude of the electric field $\boldsymbol{E}_\mathrm{LO}$ from a local oscillator (LO) at  $\omega_\olo$, so that the change in probe transmission for small changes in microwave field amplitude is maximized. Thus, any electric field $\bEA$ at frequency $\wsig$ 
superimposed with the LO, generates an 
intermediate frequency (IF) $\omega_\mathrm{IF} = \wsig-\omega_\olo$. The IF field modulates the AT splitting such that {$\Omega(t) = \Omega_0 + \Delta \Omega(t)$}, where $\Delta\Omega(t)\propto \abs{\EA}\cos(\omega_\mathrm{IF}t+\varphi(t))$. Here $\EA \ll \abs{\boldsymbol{E}_\mathrm{LO}}$ is the component of the vector $\bEA$ the atoms are sensitive to (parallel to $\boldsymbol{E}_\mathrm{LO}$), and $\varphi(t)$ is the phase of $\EA$ relative to the LO phase. This modulation is transformed into an amplitude modulation of the transmitted probe at a specific frequency 
$\nu_p$. The modulation index is maximized by choosing $\nu_p$ to lie on the maximum slope of the transmitted probe magnitude with respect to variations in $\Omega$.

	\begin{figure*}[ht]
		\centering
        \centering
		\includegraphics[width=0.95\textwidth]{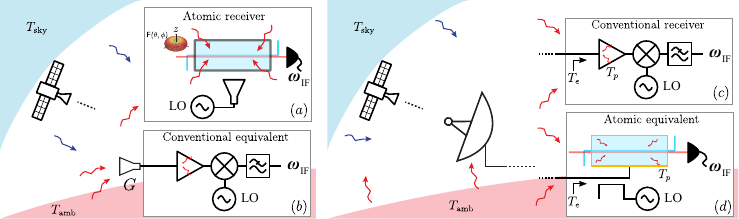}
         
		\caption{Two approaches to compare the sensitivity of atomic microwave receivers with conventional ones. On the left side, the functionality of a free-space coupled atomic receiver (a) can be replicated using an antenna-fed LNA followed by a mixer driven by an LO (b). The gain of the antenna is key in the comparison and can be chosen such that (i) its radiation (reception) pattern matches that of the atomic receiver, or (ii) it is maximized while occupying the same volume as the atomic receiver. On the right side, a port-fed conventional receiver (c) is compared with a field-enhanced atomic receiver coupled to an input port (d). The field enhancement is provided by a resonant and/or confining microwave structure. }
		\label{fig:LNA}
	\end{figure*}
	
	The presence of an RF local oscillator turns the atomic sensor into a coherent receiver. When 
	$\omega_\olo = \wsig$ the receiver is of a 
	\emph{homodyne} type 
	which is responsive only to the component of the signal that is in phase with the LO. In contrast, when the local oscillator is detuned by an intermediate frequency $\omega_\mathrm{IF}=\wsig-\omega_\olo$ the receiver becomes 
	\emph{heterodyne}, 
	where both quadrature components of the microwave signal at $\wsig$ are present in the IF signal 
	at $\omega_\mathrm{IF}$. On the same beatnote at $\omega_\mathrm{IF}$ are also superimposed both quadrature components of the \emph{image} at $\omega_i=\omega_\olo - \omega_\mathrm{IF}$. In this case, the receiver acts as a mixer or downconverter, so thermal and quantum noise present in the image band add to the total noise at the intermediate frequency. 
	
	Typically, the noise floor measured at the IF output of the atomic RF receiver shown in Fig. \ref{fig:simple}(e) has a locally flat power spectral density. Therefore, the IF noise power scales with the post-photodetection observation bandwidth $\Delta f$. Since the beatnote power is proportional to $\abs{\EA}^2$, the minimum detectable squared-field $\abs{\EA}_\mathrm{min}^2$ (i.e., that which produces a unity SNR at IF) is proportional to $\Delta f$. Equivalently, the minimum detectable electric field amplitude scales with $\sqrt{\Delta f}$. The ratio 
	
	\begin{equation}
	    \mathrm{NEF} = \frac{\abs{\EA}_\mathrm{min}}{\sqrt{\Delta f}}
	\end{equation}
	%
	is the \emph{noise equivalent field} (NEF), which is a useful sensitivity figure of merit independent from $\Delta f$.
	
	Consider the scenario illustrated in Fig. \ref{fig:LNA}(a) where a free-space-coupled atomic microwave sensor is used to receive information from a distant transmitter (e.g., a satellite or an astronomical source). Since the atoms are sensitive to properly polarized incoming waves regardless of their direction (i.e., have an omnidirectional reception pattern), in-band ambient black body radiation is received along with the signal. This thermal background determines a noise floor $\mathrm{NEF}_\mathrm{th}$ which is \emph{extrinsic} to the receiver but unavoidable in a free-space configuration. Because of the coherent nature of the sensor, the fluctuations of the vacuum microwave field $\mathrm{NEF}_q$ must be added to the extrinsic sources, i.e., $\mathrm{NEF}_\mathrm{ex}^2 = \mathrm{NEF}_\mathrm{th}^2 + \mathrm{NEF}_q^2$. The \emph{intrinsic} noise of the atomic receiver includes probe fluctuations after propagating through the atoms $\mathrm{NEF}_\mathrm{at}$, photon shot noise of the probe $\mathrm{NEF}_p$ and photodetector noise $\mathrm{NEF}_\mathrm{pd}$. Assuming these sources are uncorrelated and additive, the intrinsic noise floor of the receiver is $\mathrm{NEF}_0^2 = \mathrm{NEF}_\mathrm{at}^2 + \mathrm{NEF}_p^2+\mathrm{NEF}_\mathrm{pd}^2$, and the total NEF takes the form

	\begin{equation}
	    \mathrm{NEF} = \sqrt{\mathrm{NEF}_\mathrm{ex}^2 + \mathrm{NEF}_0^2}.\label{eq:NEFtot}
	\end{equation}

	In this work, two approaches are followed to assess the sensitivity of atomic RF receivers by comparing them with their conventional counterparts. The first approach 
	entails the definition of a conventional RF receiver (Fig. \ref{fig:LNA}(b)) that has a similar functionality to a free-space-coupled atomic receiver (Fig. \ref{fig:LNA}(a)). The second approach calls for the definition of a single-mode microwave port that feeds an atomic receiver (Fig. \ref{fig:LNA}(d)). Introducing a port enables a noise temperature to be determined in the atomic receiver, allowing for a direct comparison with the noise figure of conventional RF frontends (Fig. \ref{fig:LNA}(c)). In this scenario, microwave structures that confine and enhance the field 
	can be designed to improve the sensitivity of an atomic sensor with respect to a free-space configuration. 

	%


\subsection{Free-space-coupled receivers}
	
In the scenario shown in Fig. \ref{fig:LNA}(b) the noise performance of the receiver is dominated by the noise figure of the LNA provided that its gain is sufficiently high. Because the noise figure of an amplifier is a power spectral density metric referred to its input port, an antenna and its gain $G$ 
must be defined in order to translate noise temperature into NEF. 
A properly polarized electric field amplitude ${E}_A$ at the antenna implies an input power to the LNA equal to $G\frac{\lambda_0^2}{4\pi}\frac{\abs{\EA}^2}{2\eta_0}$ where $\lambda_0=c/f$ is the free-space wavelength, $c$ the speed of light in vacuum, $f$ the microwave signal frequency, $\eta_0=\sqrt{\mu_0/\varepsilon_0}$ the free-space impedance, and $\mu_0$ and 
$\varepsilon_0$ the vacuum permeability and permittivity respectively. The input-referred noise floor of the receiver is determined by the LNA's noise temperature $T_e = T_\mathrm{LNA}$ and is given by $k_B T_\mathrm{LNA}\Delta f$ where $k_B$ is the Boltzmann constant and $\Delta f$ the measurement bandwidth. Solving for the electric field amplitude that produces an output power equal to the noise floor we obtain 
	
	\begin{equation}
	\abs{\EA}_\mathrm{min}= \underbrace{\sqrt{\frac{8\pi f^{2}}{\varepsilon_{0}c^{3}G}}}_\Lambda  \sqrt{k_B T_\mathrm{LNA}\Delta f}
	    \label{eq:Lambda}
	\end{equation}
	Therefore, the noise in the receiver can be attributed to an input ``electric field spectral density'' or NEF 
	
	\begin{equation}
	    \mathrm{NEF}_0=
	    \Lambda\sqrt{k_B T_\mathrm{LNA}}.\label{eq:T2NEF}
	\end{equation}

Defining a unique value of $G$ that leads to a fair comparison is not trivial. For instance, if replicating the reception pattern of the atomic receiver is desired, one could take the setup in Fig. \ref{fig:LNA}(a) and measure the relative amplitude of the IF signal as a function o the incidence angle of the RF plane wave, quantifying an equivalent radiation pattern $\mathsf{F}(\theta,\phi)$ in spherical coordinates. This radiation pattern thus defines the maximum gain of a lossless antenna as $G=4\pi/\int_{0}^\pi \int_0^{2\pi}\mathsf{F}^2(\theta,\phi)\sin\theta\,\mathrm{d}\theta\mathrm{d}\phi$. If the atoms are omnidirectional but sensitive to a single vector component of the microwave field \textemdash which is determined by the polarization of the optical beams and microwave LO\textemdash then $G=3/2$.  
	However, one could also argue that a fair comparison only exists  when the conventional receiver uses an antenna whose gain is maximized while occupying the same volume as the interrogated atoms. Maximizing $G$ would indeed improve the electric field sensitivity of the LNA, forcing free-space-coupled atomic receivers to have stricter sensitivity requirements to compete with conventional RF frontends. 
	Not considering antenna superdirectivity \cite{ziolkowski2017,chu,harrington} 
	the maximum achievable gain in a spherical region of diameter $d$ can be defined as $G= (\pi d/\lambda_0)^2 + 2\pi d/\lambda_0$ \cite{harrington}. In the rest of this work, we use $G=3/2$ for comparison purposes. This is the most conservative value for conventional RF frontends, and thus, the most optimistic one for atomic receivers. It also corresponds to the minimum gain achievable by an electrically-small single-polarization lossless antenna.

	As an illustrative example, consider the state-of-the-art sensitivity of $1.25\uvmshz$ reported in \cite{cai2022} using a setup coupled to $f\sim 10\,\mathrm{GHz}$ free-space radiation. Assuming that the thermal background contribution to the noise floor of the experiment is negligible (as will be shown in Section \ref{sec:free}), let us compare this sensitivity to that achievable by an off-the-shelf room-temperature X-band LNA (Minicircuits PMA-183PLN-D+) with $T_\mathrm{LNA}\approx 100\,\mathrm{K}$ ($1.3\,\mathrm{dB}$ noise figure), coupled to an electrically small, matched and lossless antenna with gain $G=3/2$. Using Eq. (\ref{eq:T2NEF}) the LNA sensitivity is found to be $\mathrm{NEF}_0=0.098\uvmshz$, 
	about 13 times lower noise floor than that achieved by the state-of-the-art atomic sensor. In other words, an LNA with noise temperature $T_\mathrm{LNA}=16100\,\mathrm{K}$ would have the same receiver sensitivity as the best demonstrated EIT Rydberg electrometer. 
	
	\subsection{Port-coupled receivers}
	
	Now consider the scenario shown in Fig. \ref{fig:LNA}(d). 
	Using structures that confine the fields either resonantly, via mode volume reduction, or by a combination of both, can be leveraged to increase the sensitivity of the atomic receiver. Suppose an RF structure that confines the electric field sensed by the atoms is coupled to an input port carrying power $P_\mathrm{in}$. The internal electric field amplitude at the location of the interrogated atoms is $\abs{\ER}= K\sqrt{P_\mathrm{in}}$ where $K$ is a constant that depends on the structure design. 
	When this port-coupled receiver is fed with an antenna, we can define a field enhancement factor $F$, as the ratio between the external and internal field amplitudes. Similar to the result in Eq. (\ref{eq:Lambda}), 
	
	\begin{equation}
	    F = \frac{K}{\Lambda} = \frac{\abs{\ER}}{\Lambda\sqrt{P_\mathrm{in}}}.\label{eq:Fenh}
	\end{equation}
	%
	If the sensitivity of a free-space coupled atomic receiver is 
	$\mathrm{NEF}_0$, 
	the sensitivity of a field-enhanced atomic receiver $\mathrm{NEF}\approx \mathrm{NEF}_0/F$ 
	only when 	$\mathrm{NEF}_0$ is high enough for the thermal background to be negligible, or equivalently, in a $0\,\mathrm{K}$ environment. Otherwise, the thermal noise introduced by the confining structure needs to be included.  
	Even in a cryogenic environment 
	the assumption $\mathrm{NEF}=\mathrm{NEF}_0/F$ is 
	not correct because 
	$F$ can be made very large by e.g., reducing the mode volume in a superconducting resonator. 
	However, the resulting $\mathrm{NEF}$ of the receiver cannot be arbitrarily small as that would violate the quantum limit for coherent detection.  In Section \ref{sec:mode}, we develop a complete theoretical model that takes into account the effects of thermal radiation and vacuum fluctuations to field-enhanced atomic receivers. 
	We further derive trade-offs that maximize the senstivity of field-enhanced receivers leading to optimal design rules for the microwave structures. 
	
	As will be discussed in Section \ref{sec:extrap}, LNAs operating at a physical temperature $T_p$ equal to room temperature and with noise temperatures $T_e<T_p$ are readily available below W-band. Furthermore, since the LNA is port-fed, a receiver chain whose system temperature $T_\mathrm{sys}$ is independent of the ambient thermal background $T_\mathrm{amb}$ can be built with directive antennas. This independence is of particular importance in applications where the antenna is pointed to a cold background (see Fig. \ref{fig:LNA}(c)) where $T_\mathrm{sys}=T_\mathrm{sky}+T_e$ can be significantly below $T_p$ and $T_\mathrm{amb}$. This is not achievable in atomic receiver configurations as the ones shown in Fig. \ref{fig:LNA}(a), unless the microwave field sensed by the interrogated atoms is electromagnetically coupled to a single mode that defines a port as shown in Fig. \ref{fig:LNA}(d). This is another motivation to use port-fed microwave structures that confine and/or enhance the electric field.

\section{{Noise in Free-Space Receivers}}	
\label{sec:free}

	In this section, we focus on an idealized intrinsically noiseless free-space atomic receiver, and quantify the electric field sensitivity limits due to the two extrinsic noise sources: (1) black-body radiation from the surroundings at physical temperature $T$; and (2) vacuum fluctuations of the microwave field. 
	Suppose the interrogated atoms are located at $\boldsymbol{r}_0$ and are directly sensing the electric field of a plane wave propagating in free space at frequency $ \fsig =\wsig/(2\pi)$. This signal is retrieved by e.g., measuring the transmitted probe or any other band-limited measurement mechanism with single-sided power transfer function $H(f)$. The power-equivalent observation bandwidth $\Delta f = H^{-1}( \fsig )\int_0^\infty H(f)\,\mathrm{d}f$ is sufficiently narrow to assume the receiver response to be spectrally flat within $\Delta f$. 
	
	The black-body radiation emitted by the surroundings  
	of the receiver generates an electric field at $\boldsymbol{r}_0$ that adds to the signal at $ \fsig $. We assume the receiver is surrounded by media at room temperature $T$, forming an electrically large enclosure of volume $V$. This approximates the conditions encountered in e.g., a laboratory. 
	From the equipartition theorem, every mode in $V$ at a frequency $ \fsig $ has a mean thermal energy $\expval{W_\mathrm{mode, th}} = \nth h \fsig $ where $\nth$ is the boson Bose-Einstein distribution:
	
	\begin{equation}
	    n_\mathrm{th}(f,T) = \frac{1}{\exp\left(\frac{hf}{k_B T}\right)-1}.
		\label{eq:Wmode_thermal}
	\end{equation}
	It can be shown \cite{oliver} that the number of allowed modes around $f$ is proportional to $V$ and $f^3$ and tends to a continuum as $V\rightarrow \infty$. 
	The volumetric and spectral density of modes in an electrically large 3D cavity is 
	%
	$\rho_v= {8\pi f^2}/{c^3}$ 
	%
	modes per unit volume per unit bandwidth, 
	where the two polarization degrees of freedom per each spatial mode have been taken into account. Hence, the volumetric energy density of the thermally-generated electromagnetic field at any point $\boldsymbol{r}_0$, observed within a narrow bandwidth $\Delta f \ll  \fsig $ is 	\footnote{Note that Eq. (\ref{eq:WEM}) can also be derived by integrating the uniform specific brightness of the volume's walls $B_f=({2h  \fsig ^3}/{c^2})\left(\exp({h  \fsig }/{k_B T})-1\right)^{-1}$ over the $4\pi$ solid angle and bandwidth $\Delta f$ \textemdash resulting in the total surface power density\textemdash and multiplying by the propagation time per unit length $c^{-1}$ to give volumetric energy density. } $\rho_v\expval{W_\mathrm{mode,th}}\Delta f$, i.e.,
	
	\begin{equation}
		\expval{W_\mathrm{EM,th}} = \frac{8\pi h  \fsig ^3\Delta f}{c^3} \nth
		\label{eq:WEM}
	\end{equation}
    where $\expval{\cdot}$ is an expected value operator. 
	In a heterodyne setup, the actual thermal energy is twice that of Eq. (\ref{eq:WEM}) because of the downconverted thermal noise from the image band. In a homodyne setup, however, the actual thermal energy is half of Eq. (\ref{eq:WEM}) because only one quadrature of the electromagnetic field is observed and both are equally thermally populated. 
	
	On top of thermal noise, there are vacuum (zero-point) fluctuations of the microwave field that must be accounted for. In a homodyne architecture, they add half a photon per degree of freedom or mode \cite{shapiro}. In a heterodyne receiver, an extra penalty is incurred for the retrieval of both amplitude and phase information, and therefore the measurement is subject to the quantum  limit of coherent detection, i.e., one photon per degree of freedom or mode \cite{kerr, shapiro, caves}. Therefore, in each case, the sum of both thermal and quantum noise contributions to the total energy density of the electromagnetic field at $\boldsymbol{r}_0$ becomes 
	
		\begin{equation}
		\expval{W_\mathrm{EM}} = \frac{8\pi h  \fsig ^3\Delta f}{c^3} \Theta( \fsig , T),
		\label{eq:WEMtot}
	\end{equation}
	where $\Theta(f, T)$ is a modified version of the Callen-Welton law \cite{callen}:

\begin{equation}
    \Theta(f, T)=hf\begin{cases}
\frac{1}{2}n_\mathrm{th}(f,T)+\frac{1}{2} & \text{homodyne}\\
2n_\mathrm{th}(f,T)+1 & \text{heterodyne}
\end{cases}\label{eq:CallenWelton}
\end{equation}
	
	
	The instantaneous energy density $W_\mathrm{EM}(t)$ associated with the detected field is a $\chi^2$-distributed random process whose power spectral density is contained within the observation band $\Delta f$. Its mean value has equal contributions from the mean electric $\expval{W_e}$ and magnetic $\expval{W_m}$ energy densities. 
	If $\boldsymbol{E}(\boldsymbol{r}_0,t)$ is the slow-varying complex amplitude of the electric field centered at $ \fsig $ within $\Delta f$, then $\expval{W_e}=\frac{\varepsilon_0}{4}\expval{\abs{\boldsymbol{E}(\boldsymbol{r}_0,t)}^2} = \frac{1}{2}\expval{W_\mathrm{EM}}$. The components of the vector $\boldsymbol{E}(\boldsymbol{r}_0,t)$ are uncorrelated identically-distributed complex Gaussian processes with zero mean. If the atoms are sensitive to a single vector component $\hat{\boldsymbol{a}}_e$ of the field, then we can write that component  as $E=\boldsymbol{E}\cdot \hat{\boldsymbol{a}}_e$ and $\expval{\abs{\boldsymbol{E}(\boldsymbol{r}_0,t)}^2} = 3\expval{\abs{E(\boldsymbol{r}_0,t)}^2}$ because all polarization degrees of freedom are equally populated by thermal and vacuum fields. 
	Solving for the standard deviation of the observed field $\Delta E={\expval{\abs{{E}(\boldsymbol{r}_0,t)}^2}}^{1/2}$ and normalizing by the square root of the bandwidth leads to the bandwidth-independent NEF

	\begin{equation}
		\frac{\Delta E}{\sqrt{\Delta f}} = \mathrm{NEF}_\mathrm{ex}= \sqrt{ \frac{16\pi  \fsig ^2}{3\varepsilon_0 c^3}\Theta( \fsig , T)}.
		\label{eq:NEFthermal}
	\end{equation}
	%
	Note that from a receiver perspective, the atoms have an effective gain $G=3/2$ which is concluded from thermodynamic principles by comparing Eq. (\ref{eq:NEFthermal}) with eqs. (\ref{eq:T2NEF}) and (\ref{eq:Lambda}). This is a consequence of two assumptions: i) the atoms are sensitive to a single vector component of the electric field; and ii) the excited atoms are confined to an electrically-small volume where the phase difference between the microwave signal and LO is constant. 
	When the atomic interaction volume is not electrically-small, one can divide it into a discrete set of electrically small volumes with uniform signal and LO slow-varying phase difference $\varphi$. Each volume element located at $\boldsymbol{r}$ induces a small IF modulation $\propto\abs{\EA (t)}e^{i\omega_\mathrm{IF}t}e^{i\varphi(\boldsymbol{r}, t)}$ in the probe beam 
	that adds coherently to the total transmitted probe. The spatial distribution of $\varphi(\boldsymbol{r}, t)$ depends on the angle of arrival of signal and LO waves and can lead to non-omnidirectional reception patterns $\mathsf{F}(\theta,\phi)$ for the atomic ensemble. In this case, equivalent gain values $G>3/2$ can can exist, and the background-limited minimum detectable field of Fig. \ref{fig:nefplot} shall be corrected by multiplying by $\sqrt{3/(2G)}$ (see Appendix \ref{ap:F} for more details). For incoherent electrometers that do not use a microwave LO, the results of Fig. \ref{fig:nefplot} are valid for electrically large interaction volumes, as long as they are small compared to the coherence length $c/\Delta f$. 

	\begin{figure}[t]
		\centering
		\includegraphics[width = 0.49\textwidth]{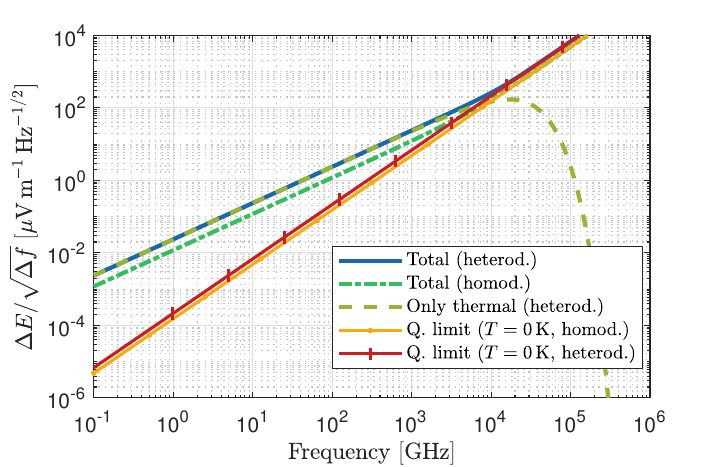}
		\caption{Noise equivalent field limited by thermal background ($T=290\,\mathrm{K}$), vacuum fluctuations of the microwave field (quantum limit for coherent detection), and both. Microwave homodyne and heterodyne detection cases are considered. Dual sideband mixer operation (DSB) is assumed for heterodyne detection. 
		}
		\label{fig:nefplot}
	\end{figure}

	Since the atoms are equally sensitive to the quantum and black-body fields and to the RF signal field, the total NEF of the system can be found using Eq. (\ref{eq:NEFtot}) 
	%
	%
	 where $\mathrm{NEF}_0$ is the NEF the atomic sensor would hypothetically exhibit in a $0\,\mathrm{K}$ background and in the absence of vacuum fluctuations. Figure \ref{fig:nefplot} shows the plot of Eq. (\ref{eq:NEFthermal}) for homodyne and heterodyne schemes at room temperature ($T=290\,\mathrm{K}$) along with the quantum limit ($T=0\,\mathrm{K}$). 
	 At room temperature, vacuum fluctuations take over thermal noise at $\approx 4.2\,\mathrm{THz}$. 
	 


\section{Noise in Port-Coupled Receivers}\label{sec:mode}
	
	Microwave cavities can be used to enhance the field the atoms are exposed to, thus improving receiver sensitivity. As illustrated in Fig. \ref{fig:cavity_sketch1}, we assume the resonator is port-coupled, and can be fed in a number of ways, including by an antenna with noise temperature $T_A$. 
	We begin by analyzing the thermal and vacuum-fluctuation contributions of both the lossy cavity and the input. While the noise originating inside the microwave cavity is now intrinsic to the atomic receiver, the input thermal background is not. Therefore, setting $T_A=0$ in the following results yields the intrinsic noise of the receiver, which remains quantum limited by virtue of Eq. (\ref{eq:CallenWelton}). 
	To find the receiver noise temperature, we first use a simple single-mode harmonic oscillator model. Then, a more general approach is followed to find the spatial distribution of thermal and vacuum fluctuations in an arbitrarily terminated waveguide. The two models agree in the high-Q limit, but the more general model extends the validity to arbitrarily overcoupled 1D resonators where an unlimited number of longitudinal modes are excited.  
	
	\subsection{Harmonic oscillator model}\label{sec:HO}
	
\begin{figure}[t]
		\centering
		\includegraphics[width = .4\textwidth]{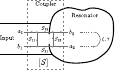}
		\caption{Sketch of a port-fed resonator. The coupling is modeled through a 2-port lossless and reciprocal network with scattering parameter matrix $\left[S\right]$. The resonator is characterized by an attenuation factor ${\satt}$ and a free spectral range which defines the round-trip delay $\tau$. }
		\label{fig:cavity_sketch1}
	\end{figure}

	Consider a resonator coupled to an input waveguide or transmission line by a lossless and reciprocal coupling structure (see Fig. \ref{fig:cavity_sketch1}). 
	%
	%
%
Its scattering parameters are defined in terms of traveling waves $a_i$, $b_i$, $i=1,2$ in lossless ports such that $\left|a_i\right|^2$ and $\left|b_i\right|^2$ equals the power entering and leaving the port $i$, respectively. The conditions above imply 
${\Sko}={\Skt}$, 
$\left|{\Sro}\right|=\left|{\Srt}\right|$, and 
$\left|{\Skt}\right|^2={1-\left|{\Srt}\right|^2}$. 
We consider the resonator to have a single round-trip time $\tau$, which corresponds to modes with a single degree of freedom, e.g. a short-circuited transmission line.  
This implies the existence of multiple resonant frequencies separated by a free spectral range ($\mathrm{FSR}$) which is constant and equal to $\tau^{-1}$ in the absence of dispersion. 
$\tau$ is defined in terms of a phase delay, in contrast to the group delay-related round-trip time $\tau_g$. 
If no dispersion exists then $\tau_g=\tau$. In the limit of high quality (Q) factors and in the vicinity of the resonance, the properties of arbitrary 3D resonant modes \textemdash having 3 degrees of freedom
\textemdash are well approximated by a harmonic oscillator (0-D cavity such as e.g., an RLC circuit).
	
Let $a_i(t)$ and $b_i(t)$ be slow-varying amplitudes of traveling waves at frequency $\omega = \omega_0 + \delta\omega$, detuned by $\delta\omega$ from the resonant frequency $\omega_0$. 
According to Fig. \ref{fig:cavity_sketch1}, a wave entering the resonator reflects with some attenuation and round-trip phase and group delays at the coupler port of: $a_2(t) = {\satt} \exp(-i\omega\tau)b_2(t-\tau_g)$, where ${\satt}$ is the round-trip field attenuation factor. 
Therefore, the reflected and intra-cavity waves take the form

\begin{align}
	b_1(t)&={\Skt} {\satt}e^{-i\omega\tau}b_2(t-\tau_g)+{\Sro}a_1(t)\label{eq:caveqb1}\\
	b_2(t)&={\Srt} {\satt}e^{-i\omega\tau}b_2(t-\tau_g) + {\Sko}a_1(t).\label{eq:caveq1}
\end{align}


If the coherence time of the intra-cavity field is much longer than the round-trip time $\tau_g$, a first order Taylor expansion suffices to approximate $b_2(t-\tau_g) \approx b_2(t) - \tau_g b_2'(t)$. When $\tau_g$ is on the same order as $\tau$, this condition implies the resonance linewidths to be much narrower than the FSR, which in turn, implies high-Q and high-finesse conditions as well as single-mode operation. The Taylor expansion transforms Eq. (\ref{eq:caveq1}) into a rate equation of the harmonic oscillator for the slow-varying intra-cavity field:

\begin{equation}
	b_2'(t)  +	\frac{e^{(\gamma + i\delta\omega)\tau} - 1}{\tau_g}{b_2(t)}{}  =     \frac{{\Sko}}{\tau_g}e^{(\gamma + i\delta\omega)\tau}a_1(t).
	\label{eq:rateeq}
\end{equation}

%
To derive Eq. (\ref{eq:rateeq}) we put ${\Srt}=e^{-\gamma_c\tau}e^{i\varphi_r}$, ${\satt}=e^{-\gamma_i\tau}e^{i\varphi_s}$ where $\gamma_c$ and $\gamma_i$ are real-valued coupling and intrinsic loss rates  in $\mathrm{rad/s}$, respectively. Then $\gamma=\gamma_c+\gamma_i$ is the total loss rate, and $\varphi_r$ and $\varphi_{\satt}$ are the phases of ${\Srt}$ and ${\satt}$, respectively. 
Furthermore, the resonance condition that ensures intra-cavity constructive interference $\omega_0\tau - \varphi_r - \varphi_{\satt} = 2m\pi\; (m\in \mathcal{Z})$ was used. Defining the coupling rates in terms of the group delay $\tau_g$ instead of $\tau$ would have been equally valid. 


To find the signal-to-noise ratio (SNR) of the receiver we proceed in the following manner: 
\begin{itemize}
    \item[(i)] The intra-cavity energy spectral density $w_\mathrm{sig}(\delta\omega)$ is calculated for a given power spectral density (PSD) of the input signal.
    \item[(ii)] The intra-cavity energy spectral density $\expval{w_n(\delta\omega)}$ due to thermal and vacuum noise generated inside the cavity, as well as coming from the background at the input, are calculated.
    \item[(iii)] Knowing the relationship between the intra-cavity energy and the electric field amplitude at the location of the Rydberg atoms, the intrinsic $\mathrm{NEF}_0$ of the atoms in a free-space configuration are referred to a stored energy spectral density $\expval{w_0(\delta\omega)}$. 
\end{itemize}
Following this procedure, the signal-to-noise ratio (SNR) is calculated as 


\begin{equation}
	\mathrm{SNR}=\frac{\expval{w_\mathrm{sig}(\delta\omega)}}{\expval{w_n(\delta\omega)}+\expval{w_0(\delta\omega)}},
\end{equation}
which is unity when the PSD of the input signal $\abs{a_1}^2$ equals the Noise Equivalent Power (NEP, in SI units $\mathrm{W\,Hz^{-1}}$). The detailed derivations of stored energy densities are available in Appendix \ref{ap:A}. Solving for the NEP, we obtain

\begin{equation}
\mathrm{NEP}	=\Theta( \fsig ,T_{A})+\frac{1}{C}\Theta( \fsig ,T_{p})+\frac{1}{K^{2}}\mathrm{NEF}_{0}^{2},\label{eq:NEP}
\end{equation}
where 
\begin{equation}
C \approx\frac{\gamma_{c}}{\gamma_{i}},\label{eq:C}
\end{equation}
is a dimensionless radiative cooling factor, and

\begin{equation}
	K^{2}
	\approx\frac{2\gamma_{c}}{\left(\gamma_{i}+\gamma_{c}\right)^{2}}\frac{\tau_g}{\tau}K_{U}^{2}(\boldsymbol{r}_{0})H_{L}(\delta\omega),\label{eq:Ksq}
\end{equation}
is the intra-cavity squared electric field per unit input power ($K$ has SI units $\mathrm{V\,m^{-1}\,W^{-1/2}}$). The term 

\begin{equation}
    H_L(\delta\omega)=\left[1+\left(\pi\frac{\delta\omega}{\Delta\omega_{r}}\right)^{2}\right]^{-1}
\end{equation}
is the Lorentzian function centered at $\omega_0$ with power-equivalent bandwidth $\Delta\omega_r =\int_{-\infty}^{\infty}H_L(\omega)\,\mathrm{d}\omega$, and 

\begin{equation}
	K_{U}(\boldsymbol{r}_{0})=\frac{\left|\boldsymbol{\Psi}(\boldsymbol{r}_{0})\cdot\hat{\boldsymbol{a}}_{e}\right|}{\sqrt{\frac{1}{2}\int_{V}\Re\left\{\varepsilon(\boldsymbol{r})\right\}\left|\boldsymbol{\Psi}(\boldsymbol{r})\right|^{2}\,\mathrm{d}V}}\label{eq:Ku},
\end{equation}
a real-valued scalar function obtained from the electric field distribution of the mode $\boldsymbol{\Psi}(\boldsymbol{r})$. 
Equation (\ref{eq:Ku}) is the ratio between the magnitude of the electric field component $\hat{\boldsymbol{a}}_{e}$ the atoms are sensitive to at $\boldsymbol{r}_0$, and the square-root of the associated mode stored energy. It is a property of the geometry $V$ and permittivity $\varepsilon$ of the cavity. From Eq. (\ref{eq:NEP}), the intrinsic noise temperature of the receiver ($T_\mathrm{noise}$ or $\mathrm{NET}$) is obtained by setting $T_A=0$ and dividing by $k_B$. Setting $T_A = 0$ keeps the input vacuum fluctuations which are measurable when characterizing the noise of a receiver using e.g., the Y-factor method \cite{kerr}. This allows for a direct comparison with the measurable noise temperature of e.g., LNAs or mixers. 


The approximations in Eqs. (\ref{eq:C}) and (\ref{eq:Ksq}) hold as $\gamma_{i,c}\ll\tau^{-1}$ which is an underlying assumption that guarantees single mode operation and has been followed throughout this section (see Appendix \ref{ap:A} for details). Note that $C$ is a factor that enables \emph{radiative cooling} when the resonator is overcoupled $\gamma_c>\gamma_i$ \cite{matskoPhysRevA.77.043812, SantamariaBotello:18, XuPhysRevLett.124.033602}. This effectively reduces the noise temperature generated by the resonator at physical temperature $T_p$, regardless of the temperature of the target $T_A$. Here, we are considering an \emph{input-referred} noise temperature which shall not be confused with that associated to the stored thermal energy. For instance, if $T_A = T_p=T$, the system is in thermal equilibrium and the stored energy of the resonant mode must have temperature $T$ regardless of coupling. However, from an SNR perspective, the input-referred system temperature equals $2\Theta(f_s, T)/k_B$ when the resonator is critically coupled ($C=1$), but halves for a strongly overcoupled resonator $C\rightarrow \infty$ (see Eq. (\ref{eq:NEP}) when $\mathrm{NEF}_0=0$).  

Recall that $K$ is the ratio between the intra-cavity electric field magnitude at the location of excited atoms, per square-root of incident power at the input port. $K^2$ is an intrinsic property of the resonator and coupling strength and scales proportionally with the loaded Q, and inversely proportionally with the mode volume. It is maximized at resonance ($\delta\omega=0$) and for critical coupling ($C=1$), as expected. Suppose an antenna of gain $G$ is connected to the input port, and is receiving a properly polarized electric field magnitude $\abs{\EA}$. 
Because $K^2\propto F^2$ (see Eq. (\ref{eq:Fenh})), the field enhancement factor provided by the cavity reduces the effect of the intrinsic noise observed in a free-space setup $\mathrm{NEF}_0$. 


For a given resonator loss rate $\gamma_i$, Eq. (\ref{eq:NEP}) is minimized for an optimal coupling rate $\gamma_c^{(\mathrm{opt})}$ given by 

\begin{equation}
	\gamma_c^{(\mathrm{opt})} = \gamma_i\sqrt{1+ \frac{2\Theta( \fsig ,T_p)}{\gamma_i \mathrm{NEF}_0^2}\frac{\tau_g}{\tau}K_{U}^{2}(\boldsymbol{r}_{0})H_{L}(\delta\omega)}.\label{eq:gammaopt}
\end{equation}

As the free-space sensitivity improves ($\mathrm{NEF}_0\rightarrow 0$), no field enhancement is needed and Eq. (\ref{eq:gammaopt}) tells us it is optimal to overcouple the resonator to reduce thermal noise due to its physical temperature $T_p$. On the other hand, poor free-space sensitivities (large $\mathrm{NEF}_0$) 
benefit from field enhancement and thus Eq. (\ref{eq:gammaopt}) suggests the resonator must be critically coupled ($\gamma_c=\gamma_i$). In general, Eq. (\ref{eq:gammaopt}) represents the trade-off between field enhancement and radiative cooling for minimum total noise. When an overcoupled solution is optimal, from Eq. (\ref{eq:caveq1}), the reflection coefficient of the input port becomes

\begin{equation}
    \abs{\Gamma_\mathrm{in}}=\frac{\abs{b_1}}{\abs{a_1}}\approx\abs{\frac{\gamma_c-\gamma_i}{\gamma_c +\gamma_i}}.
\end{equation}
Despite 
signal power loss due to impedance mismatch between input port and resonator, the noise and SNR performance are still optimal.

\subsection{Waveguide/resonator model}\label{sec:wg}

In the previous section 
a harmonic oscillator model is used to describe the NEF of a cavity-enhanced atomic sensor. It is found that for sufficiently low values of free-space $\mathrm{NEF}_0$, overcoupled cavities are optimal because of radiative cooling. However, since the model relies upon high-Q single mode assumptions, it is inaccurate when cavities are strongly overcoupled, or very lossy. Here we focus our attention to 1D waveguides/resonators, 
and find the thermal contributions using 
an arbitrary number of longitudinal modes. This allows arbitrary loss and coupling strengths to be accurately modeled. 

\begin{figure}[t]
	\centering
	\includegraphics[width = .4\textwidth]{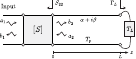}
	\caption{Sketch of 1D waveguide resonator model. The input port feeds the resonator through the 2-port coupling network $\left[S\right]$. The resonator is a waveguide of length $L$ and propagation constant $\kappa_z = \alpha+i\beta$ at physical temperature $T_p$. The waveguide is terminated in a load with reflection coefficient $\Gamma_L$ at temperature $T_L$.}
	\label{fig:wgres1}
\end{figure}

As depicted in Fig. \ref{fig:wgres1}, the resonator consists of a section of waveguide of length $L$ and complex propagation constant $\kappa_z=\alpha + i\beta$, terminated in a load whose complex reflection coefficient is $\Gamma_{L}$. It is coupled via a 2-port network described with a generalized S-parameter matrix. It is assumed that the waveguide and the terminating load are at physical temperatures $T_p$ and $T_L$, respectively. Therefore, the termination can be thought of as a mismatched antenna with reflection coefficient $\Gamma_{L}$ receiving 
black-body radiation at temperature $T_L$.

If the traveling waves $a_{1,2}$ and $b_{1,2}$ are power-normalized, then their squared magnitudes are power spectral densities dependent of $\omega$. The frequency-domain equations governing the traveling waves are

\begin{align}
	a_2 &= \Gamma_{L} e^{-2\kappa_z L}b_2 \\ 
	b_2 &= {\Srt} a_2 + {\Skt} a_1
\end{align}

Not surprisingly, the signal-related intra-cavity power has the same 
form as the steady-state detuned solutions of previous section (see Appendix \ref{ap:A}):

\begin{equation}
	b_2(\omega) = \frac{{\Skt}}{1-{\Srt}\Gamma_{L}e^{-2\kappa_z L}}.
\end{equation}
Referring to the coordinate system of Fig. \ref{fig:wgres1}, the signal-related mode distribution along the resonator is given by $b_2(\omega)e^{-\kappa_z z}\left(1+\Gamma_{L}e^{-2\kappa_z (L-z)}\right)$, whose squared-magnitude is

\begin{equation}
	\Ssig(\omega,z) = \frac{\left(1-\abs{{\Srt}}^2\right)\abs{1+\Gamma_{L}e^{-2\kappa_z (L-z)}}^2}{e^{2\alpha z}\abs{1-{\Srt}\Gamma_{L}e^{-2\kappa_z L}}^2}\abs{a_1(\omega)}^2.\label{eq:Ssig}
\end{equation}
The PSD distribution of Eq. (\ref{eq:Ssig}) must be compared with the thermal noise-related PSD distribution to find a spatial-dependent SNR. Suppose the input of the resonator is coupled to a matched antenna whose radiation pattern is surrounded by a thermal background at temperature $T_A$. This source of noise produces a similar PSD distribution 

\begin{equation}
	\Sa(\omega,z) = \frac{\left(1-\abs{{\Srt}}^2\right)\abs{1+\Gamma_{L}e^{-2\kappa_z (L-z)}}^2}{e^{2\alpha z}\abs{1-{\Srt}\Gamma_{L}e^{-2\kappa_z L}}^2}{\Theta( \fsig ,T_A)}{}.\label{eq:SA}
\end{equation}
If the waveguide termination is not perfectly reflective $\abs{\Gamma_{L}}<1$, and has 
temperature $T_L$, 
it will also cause a thermal distribution $\Sl(\omega,z)$. By symmetry, the expression for $\Sl(\omega,z)$ follows that of Eq. (\ref{eq:SA}) after replacing $z\rightarrow L-z$, ${\Srt}\rightarrow \Gamma_{L}$, $\Gamma_{L}\rightarrow {\Srt}$ and $T_A\rightarrow T_L$.

\begin{figure}[t]
	\centering
	\includegraphics[width = .45\textwidth]{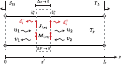}
	\caption{Illustration of the impressed equivalent sources at $z=z'$ that generate uncorrelated thermal noise waves $d_\pm^i$. These waves interact with the boundary conditions of the resonator $\Gamma_L$ and ${\Srt}$, yielding $u_{1,2}$, and $v_{1,2}$ traveling waves at $z=z'$.}
	\label{fig:JiMi}
\end{figure}

The second source of thermal and quantum noise is the lossy waveguide itself. To evaluate the spatial profile of noise 
in an arbitrary electromagnetic structure one can find appropriate zero-mean, Gaussian-distributed, white and uncorrelated impressed sources $\boldsymbol{J}_i(\boldsymbol{r}, t)$ that generate thermal radiation. Such sources are spatially distributed, with a cross-correlation given by the dyadic \cite{haus}

\begin{multline}
	\expval{\boldsymbol{J}_i(\boldsymbol{r}, t)\boldsymbol{J}_i^{*}(\boldsymbol{r}', t')} = 2\Theta( \fsig ,T_p)\overline{I} \sigma(\boldsymbol{r})  \\ \times \delta(\boldsymbol{r}-\boldsymbol{r}')\delta(t-t'),\label{eq:Jhaus}
\end{multline}
where $\overline{I}$ is the unit dyadic, $\delta(x)$ the Dirac delta, $\sigma$ the conductivity of the medium, and $T_p$ the physical temperature of the structure. Integrating the stochastic sources $\boldsymbol{J}_i(\boldsymbol{r}, t)$ in Maxwell's equations is in general not possible except for simple structures. 
In the case of a metallic waveguide, one would need to integrate a distribution of impressed currents on the walls, or at all points with finite conductivity. However, our particular problem can be simplified by virtue of  the single degree of freedom of wave propagation together with thermodynamic principles. Suppose at $z=z'$, one takes a small section of waveguide of length $\Delta z$ and volume $\Delta V$ and computes the EM thermal power flow through the boundaries of $\Delta V$ due to $\boldsymbol{J}_i(\boldsymbol{r}, t)$. Since power only flows through the cross-section of the waveguide, one could replace the surfaces $S_T^+$ and $S_T^-$ (see Fig. \ref{fig:JiMi}) by sheets of equivalent electric and magnetic impressed currents $\boldsymbol{J}_\mathrm{i,eq}$, $\boldsymbol{M}_\mathrm{i,eq}$. Those currents will generate uncorrelated \emph{impressed} traveling waves $ \diplus $ for $z>z'$  and $ \diminus $ for $z<z'$.

Suppose both ends of the waveguides are terminated in matched loads, e.g. antennas, pointing to extended black bodies at temperature $T_p$. The leftmost antenna couples a PSD equal to $\Theta( \fsig ,T_p)$ to the waveguide mode, which is attenuated to $e^{-2\alpha\Delta z}\Theta( \fsig ,T_p)$ as it propagates from $z'$ to $z'+\Delta z$. To keep thermodynamic equilibrium so that the rightmost antenna radiates $\Theta( \fsig ,T_p)$ (i.e., the same it receives) the impressed wave $ \diplus $ must have a PSD equal to 

\begin{multline}
	\abs{ \diplus }^2= (1-e^{-2\alpha\Delta z})\Theta( \fsig ,T_p) 
	\\ \rightarrow 2\alpha\Theta( \fsig ,T_p)\Delta z \;\;\text{as } \Delta z\rightarrow 0.
\end{multline}
The same argument applies to the backward propagating wave, so $\abs{ \diminus }^2=\abs{ \diplus }^2$. The white spectrum $\abs{ \diplusminus }^2$ of the source at $z'$ of elementary length $\mathrm{d}z'$ produces a space and frequency dependent PSD along the waveguide 
given by

\begin{equation}
	\mathrm{d}{\Sp}(z,z') = {2\alpha\Theta( \fsig ,T_p)}{}\left[\abs{f^+(z,z')}^2 + \abs{f^-(z,z')}^2 \right]\,\mathrm{d}z'. \label{eq:dSpzzp}
\end{equation}
where $f^\pm(z,z')$ is given by Eq. (\ref{eq:fplus}) in Appendix \ref{ap:B} along with the detailed derivation of Eq. (\ref{eq:dSpzzp}).

The total spatial distribution of the PSD is obtained via integration of all uncorrelated sources ${\Sp}(\omega, z)=\int_{z'=0}^{z'=L}\,\mathrm{d}{\Sp}(z,z')$, yielding

\begin{equation}
	{\Sp}(\omega, z)=\frac{\Theta( \fsig ,T_p)}{\left|1-{\Srt}\Gamma_{L}e^{-2\kappa_{z}L}\right|^{2}}f(z),\label{eq:Sp}
\end{equation}
where

%
%

\begin{equation}
	\begin{split}
f(z) & =\left|e^{-\kappa_{z}z}+\Gamma_{L}e^{-2\kappa_{z}L}e^{\kappa_{z}z}\right|^{2}\\
 & \times\left[e^{2\alpha z}-1-\left|{\Srt}\right|^{2}\left(e^{-2\alpha z}-1\right)\right]\\
 & -\left|e^{\kappa_{z}z}+{\Srt}e^{-\kappa_{z}z}\right|^{2}\\
 & \times\left[e^{-2\alpha L}-e^{-2\alpha z}-\left|\Gamma_{L}\right|^{2}e^{-4\alpha L}\left(e^{2\alpha L}-e^{2\alpha z}\right)\right].
	\end{split}
\end{equation}
The function $f(z)$ can be thought of as a \emph{noise temperature profile} of the mode due to $T_p$ and the vacuum field. 

The last source of fluctuations is the intrinsic noise of the sensor as measured in free-space,  $\mathrm{NEF}_0$. To refer it to the propagating modes of the waveguide, we take the transversal traveling wave field distribution $\boldsymbol{\Phi}(\boldsymbol{r})$ which does not depend on $z$. 
Then, we define a normalized field magnitude at the excited atoms' location, in the direction of the observation component $\hat{\boldsymbol{a}}_e$:

\begin{equation}
K_{W}(\boldsymbol{r}_0)=\frac{\left|\boldsymbol{\Phi}(\boldsymbol{r}_0)\cdot\hat{\boldsymbol{a}}_{e}\right|}{\sqrt{\int_{S_{T}}\Re\left\{\left.\boldsymbol{S}\right|_{z=0}\right\}\,\cdot\mathrm{d}\boldsymbol{S}}},\label{eq:KW}
\end{equation}
where $\boldsymbol{S}=\left(i2\omega\mu(\boldsymbol{r})\right)^{-1}\boldsymbol{\Phi}\times\nabla\times\left(\boldsymbol{\Phi}e^{-\kappa z}\right)^{*}$ is the Poynting vector, and $S_T$ is the waveguide cross section. Squared-magnitudes of $a$ and $b$ traveling waves are transformed into squared electric field magnitudes at $\boldsymbol{r}_{0}$ (along the desired vector component) by multiplying them by $K_{W}^2(\boldsymbol{r}_{0})$. Therefore, the SNR of the system is

\begin{equation}
	\mathrm{SNR}=\frac{\Ssig}{\Sa+\Sl+{\Sp}+\frac{\mathrm{NEF}_{0}^{2}}{ K_{W}^{2}(\boldsymbol{r}_{0})}}.
\end{equation}
Using Eqs. (\ref{eq:Ssig}), (\ref{eq:SA}) and (\ref{eq:Sp}), yields a noise-equivalent power 

\begin{equation}
	\mathrm{NEP}=\Theta( \fsig ,T_{A})+\frac{\Theta( \fsig ,T_{L})}{B'}+\frac{\Theta( \fsig ,T_{p})}{C'}+\frac{\mathrm{NEF}_{0}^{2}}{K'^{2}},\label{eq:NEPwgmodel}
\end{equation}
where 

\begin{equation}
B'=\frac{e^{-4\alpha z}\left(1-\abs{{\Srt}}^{2}\right)\abs{1+\Gamma_{L}e^{-2\kappa_{z}(L-z)}}^{2}}{e^{-2\alpha L}\left(1-\abs{\Gamma_{L}}^{2}\right)\abs{1+{\Srt}e^{-2\kappa_{z}z}}^{2}},
\end{equation}

\begin{equation}
C'=\frac{e^{-2\alpha z}\left(1-\abs{{\Srt}}^{2}\right)^{2}\abs{1+\Gamma_{L}e^{-2\kappa_{z}(L-z)}}^{2}}{f(z)},
\end{equation}
and

\begin{equation}
K'^{2}=\frac{e^{-2\alpha z}\left(1-\abs{{\Srt}}^{2}\right)^{2}\abs{1+\Gamma_{L}e^{-2\kappa_{z}(L-z)}}^{2}}{\abs{1-{\Srt}\Gamma_{L}e^{-2\kappa_{z}L}}^{2}}K_{W}^{2}(\boldsymbol{r}_{0}).
\end{equation}

While $C'$ is a factor that can provide radiative cooling, $K'$ represents the magnitude of the desired component of the electric field at the location of the excited atoms per unit square-root of power at the input port of the resonator. $B'$ quantifies the suppression of noise leaked via absorption or radiation of the termination and $\rightarrow\infty$ as $\abs{\Gamma_{L}}\rightarrow1$. When the coupled waveguide is a high-Q resonator and $\abs{\Gamma_L}=1$, eqs. (\ref{eq:NEPwgmodel}) and (\ref{eq:NEP}) agree numerically. Hence, the simple harmonic oscillator model provides validation for the more general waveguide model derived in this section.

\section{Numerical results for practical cases}
\label{sec:results}

The above theoretical derivations are quite general. In this section, several practical examples are evaluated numerically. First an X-band (8.2--12.4\,GHz) metallic cavity implemented in standard WR-90 rectangular waveguide is evaluated as a field enhancement structure, using both the harmonic oscillator and waveguide models. The trade-off between coupling strength and field enhancement is evidenced. Then, the state-of-the-art sensitivity value of atomic sensors in a free-space configuration is extrapolated to other frequency bands, and compared with the noise temperature of LNAs reported in the literature.

\subsection{X-band resonator}

\begin{figure}[t]
	\centering
	\includegraphics[width = .49\textwidth]{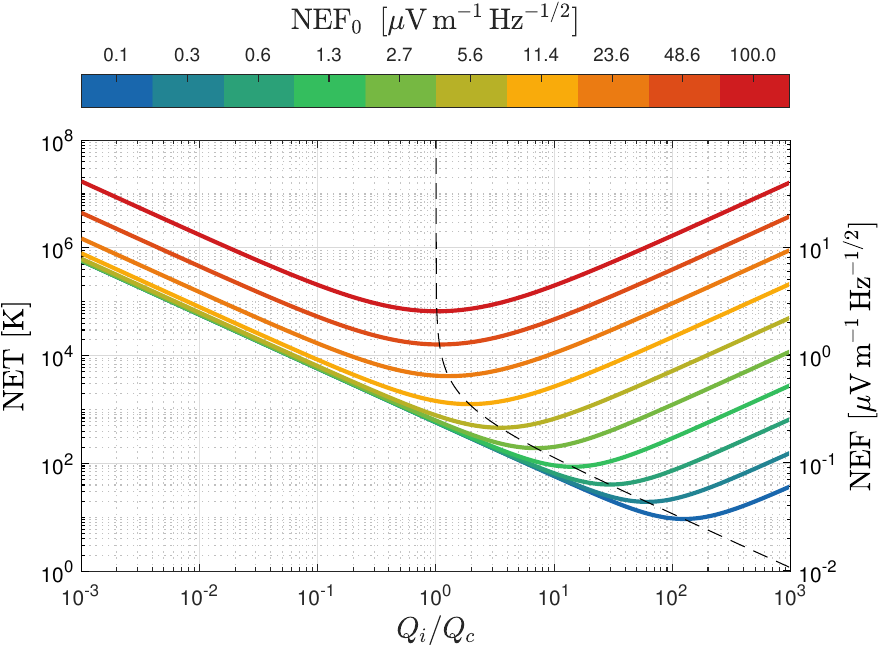}
	\caption{Resulting NET and NEF as a function of the coupling strength for different free-space $\mathrm{NEF_0}$ values in an atomic receiver enhanced by a WR90 half-wavelength waveguide resonator. The dashed line follows the minimum-noise/optimal-coupling points. }
	\label{fig:optimal_gammac}
\end{figure}

\begin{figure}[t]
	\centering
	\includegraphics[width = .45\textwidth]{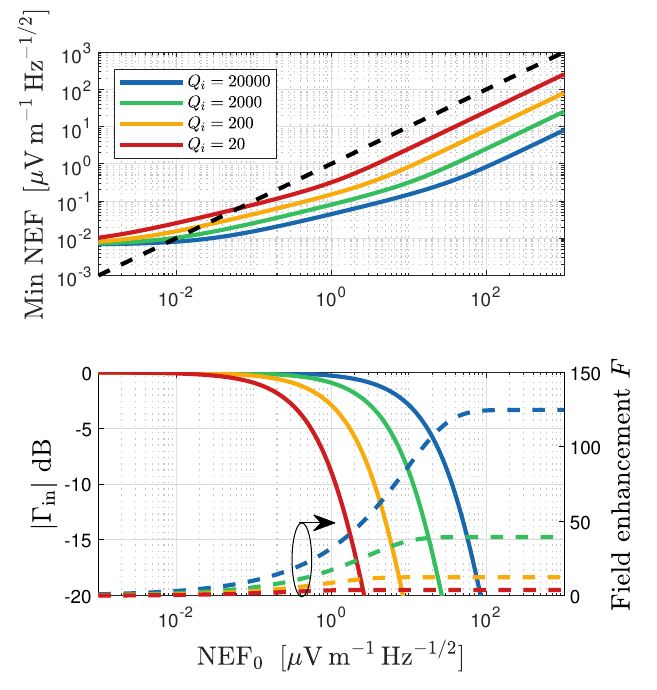}
	\caption{Top: Optimal NEF (solid lines) and $\mathrm{NEF}=\mathrm{NEF}_0$ break-even line (dashed line). Bottom: Optimal return losses (solid lines) and field enhancement factor $F$ (dashed lines). Both figures are plotted as a function of the free-space $\mathrm{NEF}_0$ and for different intrinsic $Q_i$ values in the WR90 field-enhancing resonator example.}
	\label{fig:nefnef}
\end{figure}


Consider the particular case of a room-temperature ($T_p=290\,\mathrm{K}$) half-wavelength long WR-90 waveguide resonator used to enhance the fields sensed by atoms. 
%
%
The resonant frequency is chosen to be $\wsig =2\pi\times 10\,\mathrm{GHz}$ and the intrinsic quality factor of the cavity is assumed to be $Q_i=2000$,  which is achievable with commercially available waveguides. At the geometrical center of the resonator, $K_U\approx 4.43\times10^8\;\mathrm{V\, m^{-1}\,J^{-1/2}}$ resulting in a field enhancement factor $F\approx 39.5$ at resonance and critical coupling (for a $G=3/2$ antenna). 

\subsubsection{Harmonic oscillator model}

Focusing our attention on a single resonant high-Q mode, the harmonic oscillator model is valid.  The tradeoff between field enhancement and radiative cooling is demonstrated in Fig. \ref{fig:optimal_gammac}. Color curves 
represent the Noise Equivalent Temperature (NET or $T_\mathrm{noise}$) which is obtained by dividing Eq. (\ref{eq:NEP}) by $k_B$, as a function of the ratio between intrinsic and coupling Q factors $Q_i/Q_c=\gamma_c/\gamma_i=C$ for different free-space $\mathrm{NEP_0}$ values. It can be seen how higher overcoupling rates ($Q_i/Q_c>1$) are needed as $\mathrm{NEF_0}$ decreases to realize the optimal NET. This implies the resonator is being radiatively cooled at the expense of reduced field enhancement factor $F$. The right axis shows the $\mathrm{NEF}=\varLambda \sqrt{k_B\mathrm{NET}}$, assuming an antenna with gain $G=3/2$. This plot quantifies the relationship between NEF and NET (directly related to noise figure) of an atomic receiver.  For sufficiently high values of $\mathrm{NEF_0}$, it is optimal to critically couple the resonator. The trend of the optimal coupling (Eq. (\ref{eq:gammaopt})) is illustrated in the dashed curve in Fig. \ref{fig:optimal_gammac}. 

Figure \ref{fig:nefnef} (top) shows the minimum (optimal) NEF as a function of the free-space $\mathrm{NEF}_0$ for the same WR-90 resonator example. To illustrate the effect of the intrinsic losses, curves are generated for different intrinsic $Q_i$ values.  The points below the $\mathrm{NEF}= \mathrm{NEF}_0$ dashed line belong to the region where the cavity improves the free-space behavior.  
For this particular example with $Q_i= 2000$, the break-even point occurs at $\mathrm{NEF}_0 \approx 6\,\mathrm{nV\,m^ {-1}\,Hz^{-1/2}}$, meaning that \textemdash provided optimal coupling is accomplished\textemdash introducing a WR-90 resonator with $Q_i\approx 2000$ improves the sensitivity of the atoms as long as the free-space noise equivalent field is above the break-even point. An $\mathrm {NEF_0}$ smaller than this value is below the quantum limit, (see Fig. \ref{fig:nefplot}), so in this example, the field enhancement cavity will always help.  Naturally, the break-even point decreases as the intrinsic $Q_i$ of the resonator increases. As $\mathrm{NEF}_0\rightarrow\infty$, the sensitivity improvement equals the maximum $F\approx 39.5$. The reflection coefficient $\abs{\Gamma_\mathrm{in}}$ and field enhancement factor $F$ achieved at the optimal point is also plotted for different $Q_i$ values in Fig. \ref{fig:nefnef} (bottom).

From the results above it can be seen that for the state-of-the-art $\mathrm{NEF}_0=1.25\,\uvmshz$ \cite{cai2022} the minimum noise temperature $\mathrm{NET}_\mathrm{min}\approx 87\,\mathrm{K}$ is achieved for a $Q_i/Q_c\approx 14.3$ overcoupled cavity. If the cavity is instead designed for impedance matching (critical coupling, or $Q_i/Q_c=1$) the resulting noise temperature becomes $\mathrm{NET}\approx 591\,\mathrm{K}$ which is about two times the physical temperature $T_p$ of the resonator. The factor of two arises because the receiver is effectively a mixer, coupling the resonator's thermal noise at signal and image bands. As expected from Eq. (\ref{eq:NEP}), when the effect of $\mathrm{NEF}_0$ and vacuum fluctuations becomes negligible, the noise temperature of the receiver approaches twice the physical temperature of the resonator when critically coupled. A fair comparison with LNAs would imply halving the resulting $\mathrm{NET}$ values because LNAs are not sensitive to an \emph{image band}. In other words, one could transmit the information to both signal and image bands of the mixer, which doubles the SNR and effectively halves the noise temperature. Accordingly, the optimally designed resonator would exhibit a noise temperature comparable to an LNA of about $43\,\mathrm{K}$, or equivalently, a noise figure of $\mathrm{NF}\approx 0.6\,\mathrm{dB}$. 

\begin{figure}[ht]
		\centering
        \subfloat[\label{fig:lda2}]{%
        \centering
		\includegraphics[width=0.23\textwidth]{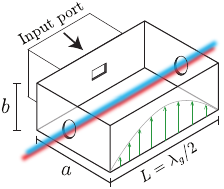}
		}
        \hfill
		\centering
        \subfloat[\label{fig:ldainf}]{%
        \centering
		\includegraphics[width=0.23\textwidth]{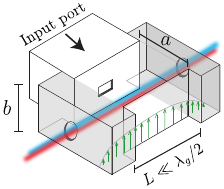}
		}
		\caption{(a) Rectangular waveguide resonator excited with the fundamental mode (half-wavelength long). (b) Rectangular waveguide resonator composed of a near-cutoff middle section for field uniformity. The two ends of the middle section (shaded region) are near quarter wavelength long and filled with glass to prevent atom excitation in the non-uniform part of the field. }
		\label{fig:wgcutoff}
	\end{figure}

{Replicating the free-space sensitivity of \cite{cai2022} in a setup involving a WR-90 resonator inserted in a vapor cell requires a similar interaction volume. The length of the WR-90 resonator is $L=17.85\,\mathrm{mm}$ while the interaction length in \cite{cai2022} is around $70\,\mathrm{mm}$. As shown in Appendix \ref{ap:C}, the difference in interaction volume can be compensated by proportionally increasing the beam waists from $\sim 1\,\mathrm{mm}$ to $\sim 4\,\mathrm{mm}$ which perfectly fits the WR-90 cross-section. The fact that the microwave field has a sine distribution in the beam propagation direction, might need to be compensated as well using slightly wider beam waists. The intrinsic Q of the resonator is almost invariant with its length because wall loss and volume grow in the same proportion. Therefore, $K^2\propto 1/L$ because energy is distributed throughout the resonator length (see Eq. (\ref{eq:Ksq})), and the ratio $\mathrm{NEF}_0^2/K^2\propto 1/L$ in Eq. (\ref{eq:NEP}). Hence, although longer cavities have lower field enhancement factors $F$, they achieve better noise performance because $\mathrm{NEF}_0$ improves faster than $F$ with respect to $L$. 
One possible design that improves field uniformity in a waveguide resonator is to terminate a near-cutoff waveguide with near quarter wavelength sections filled with glass, allowing the atoms to sense an electric field with better uniformity. Such an example is shown in Fig. \ref{fig:wgcutoff} as an illustration. }

\subsubsection{Comparison with the waveguide model}

\begin{figure}[t]
	\centering
	\includegraphics[width = .48\textwidth]{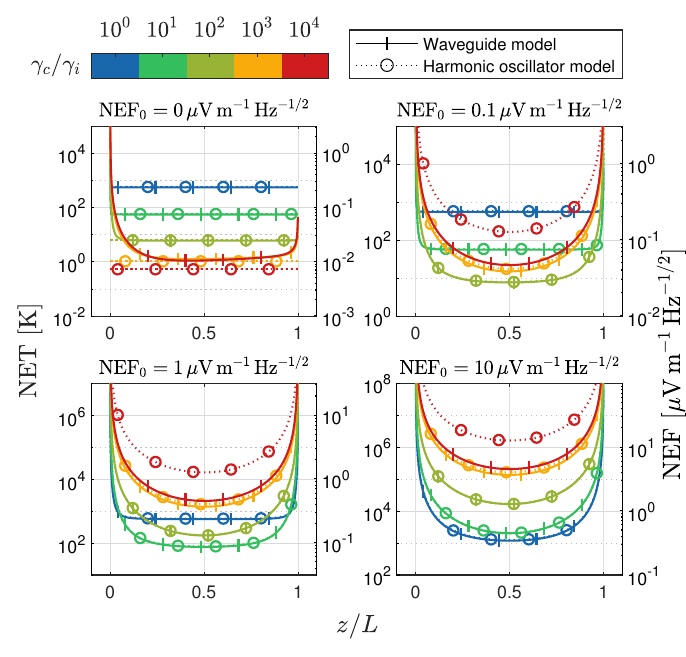}
	\caption{{Spatially-dependent NET and NEF in the WR90 waveguide resonator example calculated using the harmonic oscillator and waveguide models. Different overcoupling strengths $\gamma_c/\gamma_i=Q_i/Q_c$ are plotted for different values of free-space sensitivity $\mathrm{NEF_0}$. It can be seen how models disagree for sufficiently high overcoupling strengths. While the waveguide model is accurate in this case, the high-Q assumption the harmonic oscillator model relies on is broken. }}
	\label{fig:spatial}
\end{figure}

Let us now use the waveguide model for the example discussed above, with a  WR-90 waveguide   
terminated in a perfect reflector, $\Gamma_{L}=-1$. At the center of the cross section of the waveguide, $K_W\approx 2.931\times 10^3\,\mathrm{V\,m^{-1}\,W^{-1/2}}$.  For different coupling strengths $\gamma_c/\gamma_i$, the resulting NET (or NEF in the right axis for an antenna with gain $G=3/2$) is plotted in Fig. \ref{fig:spatial} for the two models, as a function of the position $z$ of the excited atoms along the waveguide and for several values of $\mathrm{NEF}_0$. 
When only thermal noise due to waveguide loss exists in the system ($\mathrm{NEF}_0=0$) overcoupling is expected to improve sensitivity. However, the harmonic oscillator model predicts that noise can be arbitrarily reduced down to the quantum limit by overcoupling the resonator $\gamma_c/\gamma_i\rightarrow\infty$. The waveguide model is more realistic in this scenario and accounts for the thermal noise acquired by the signal in a single round-trip as $\gamma_c/\gamma_i\rightarrow\infty$, showing a fundamental lower limit on radiative cooling. Moreover, it provides spatial information, showing the excited atoms should be located at $z\approx 0.35 L$ to minimize noise in this particular example. 
For limited coupling strengths $\gamma_c/\gamma_i \lesssim 100$, there is excellent agreement between the two models. 

Even though the harmonic oscillator model does not provide spatial information for different coupling rates, the value of $K_U(\boldsymbol{r}_0)\propto \sin(\beta z)$ can be sampled along the waveguide to estimate the spatial-dependent $\mathrm{NET}$ when $\mathrm{NEF}_0>0$ and compare with the waveguide model. 
As expected from the behavior in Fig. \ref{fig:optimal_gammac}, the minimum $\mathrm{NET}$ values are accomplished with stronger overcoupling regimes as $\mathrm{NEF}_0$ decreases. Although not plotted, both models also agree well in undercoupled regimes. Undercoupled modes are however not desired as they always exhibit more noise than overcoupled or critically-coupled ones.

	\subsection{Comparison with existing LNAs across frequency}\label{sec:extrap}
	
	In this section, the noise temperature of field-enhanced Rydberg electrometers is calculated. We use half-wavelength-long ($L=\lambda_g/2$) rectangular waveguides near cutoff as enhancement structure. 
	The longer side of the waveguide cross section $a$ is chosen so that the guided wavelength $\lambda_g=2.5\lambda_0$, where $\lambda_0$ is the free-space wavelength. The shorter side of the cross section $b=0.75a$ is chosen to allow room for wide beams while avoiding higher order modes. 
	
	\subsubsection{Extrapolation of state-of-the-art sensitivities}
	
	The experimental sensitivity results of \cite{cai2022} at $10.68\,\mathrm{GHz}$ using a $70S_{1/2}\rightarrow 70P_{3/2}$ Rb transition are extrapolated to other $nS_{1/2}\rightarrow nP_{3/2}$ Rb transitions ranging from $700\,\mathrm{MHz}$ to $298\,\mathrm{GHz}$. The extrapolation takes into account the change in three frequency-dependent factors. First, 
	since the free-space sensitivity is expected to be inversely proportional to the modulation index $m_p$, which is in turn proportional to the dipole moment $\mu_d(n)$, then $\mathrm{NEF}_0\propto 1/\mu_d(n)$ (see Appendix \ref{ap:D} and Fig. \ref{fig:dipole} \cite{SIBALIC2017319}). Second, the interaction volume shrinks as frequency increases due to the frequency-dependent size of the cavity. As shown in Appendix \ref{ap:C}, $\mathrm{NEF}_0\propto 1/(Lw_0)$, where $w_0$ is the probe beam waist (radius). In the extrapolation, we set $w_0 = \min(1.43\,\mathrm{mm}, 0.25b)$. This allows us to fit the beam in the cross-section of the waveguide, while keeping the beam reasonably small, leading to practical coupling beam powers for a given Rabi frequency. Finally, the fact that the field is sine-distributed along the cavity length is accounted for by averaging the sensed electric field. {This reduces the sensitivity by a factor $2/\pi = \pi^{-1}\int_0^\pi \sin(z)\,\mathrm{d}z$. 
	
	The non-uniformity of the field can also cause broadening of the EIT signature. Accounting for this effect would require a more detailed atomic physics modeling \cite{Hollowaybroadening}, which is beyond the scope of this work. Nevertheless, the field uniformity can be improved by using structures as the one shown in Fig. \ref{fig:ldainf}. } 	The free-space sensitivity achievable in an interaction volume determined by $w_0(n)$ and $L(n)$ is obtained by extrapolating the experimental sensitivity $n=70$ of \cite{cai2022} as

	\begin{equation}
	    \mathrm{NEF}_0 (n) = \mathrm{NEF}_0(70)\times \frac{\mu_d(70)}{\mu_d(n)}\times \frac{L(70)w_0(70)}{L(n)w_0(n)} \times \frac{\pi}{2}.\label{eq:NEF0extrap}
	\end{equation}
	The values of $w_0(n)$ and $L(n)$ are those that fit a half-wavelength cavity with the characteristics described above, resonating at the frequency of the $nS_{1/2}\rightarrow nP_{3/2}$ transition. 
	In the experiment of \cite{cai2022}, $L(70)\approx 70\,\mathrm{mm}$ and $w_0(70)\approx 0.5\,\mathrm{mm}$. However, to compensate for the shorter cavity length $L=24.56\,\mathrm{mm}$ at $10.68\,\mathrm{GHz}$, we use $w_0=1.43\,\mathrm{mm}$, which leaves the product $L(70)w_0(70)$ unchanged.

\begin{figure}[t]
	\centering
	\includegraphics[width = .43\textwidth]{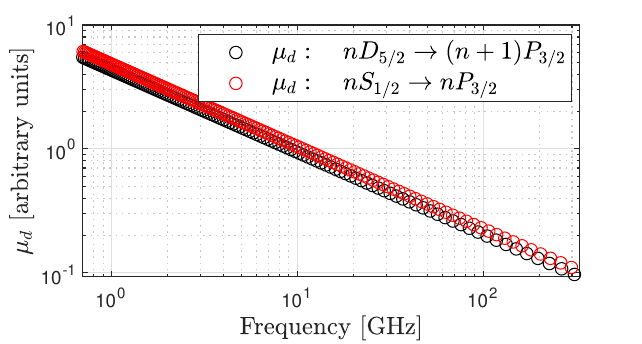}
	\caption{Dipole moment values in $^{85}$Rb as a function of the microwave transition frequency \cite{SIBALIC2017319}.}
	\label{fig:dipole}
\end{figure}
	
	Figure \ref{fig:compa} shows a survey of noise temperatures of LNAs in different technologies from $0.6\, \mathrm{GHz}\sim 329\, \mathrm{GHz}$ reported in the literature  \cite{belostotski}. Such noise temperatures are compared to those achievable in a free-space atomic setup having the interaction volume fixed to that of \cite{cai2022}, i.e., extrapolating $\mathrm{NEF}_0 (n) = \mathrm{NEF}_0(70)\times {\mu_d(70)}/{\mu_d(n)}$. The noise temperature of field-enhanced atomic receivers using optimally coupled resonators described above is also plotted. A bulk conductivity of the metallic walls  $\sigma = 25\times 10^6\,\mathrm{S/m}$ is assumed, corresponding to Aluminum 5051.  
	The calculations are done using the waveguide model \footnote{The waveguide model agrees with the harmonic oscillator model within $0.00005\%\sim 0.5\%$ relative error. As expected, the highest error is observed with the most overcoupled results. In the calculations of Fig. \ref{fig:compa10x} the maximum error between models is $6\%$.} (Eq. (\ref{eq:NEPwgmodel})) and extrapolating the non-enhanced sensitivity for the new cavity volume using Eq. (\ref{eq:NEF0extrap}). The cavity-enhanced system shows a significant improvement over a free-space setup up to W-band. The reason why using the microwave resonators is detrimental above W-band is due to the reduced interaction volume that comes from restricting the resonator length to half wavelength. To illustrate this, we have also plotted the equivalent noise temperature of a free-space setup whose interaction volume is variable, so that it fits the corresponding waveguide dimensions. This is done via Eq. (\ref{eq:NEF0extrap}) without the $\pi/2$ degrading factor. In this case, it is evident that the microwave resonator always improves a free-space configuration with the same interaction length. The knee (change in slope) observed around $23\,\mathrm{GHz}$ is a consequence of restricting the beam waist radius to a maximum of $1.43\,\mathrm{mm}$, chosen to compensate for the shorter interaction length of the cavity with respect to the experiment in \cite{cai2022}. 
	
	As frequency decreases from the upper bound, not only the length $L$ but also the cross section of the waveguides increases, leaving room for wider beams which improves sensitivity. Below $23\,\mathrm{GHz}$, the beam size is fixed \textemdash as wider beams would be impractical\textemdash leading to a slower improvement of the sensitivity with the increase in $L$. It is remarkable that for a fixed interaction volume, a free-space atomic receiver equivalent noise temperature improves at higher frequencies. The reason for this is the following. On one hand, the electric field sensitivity is degraded by the factor $\mu_d(70)/\mu_d(n)$ due to smaller dipole moments as frequency increases. This degradation is sub-linear with frequency (see Fig. \ref{fig:dipole}) which means $\mu_d \propto f^k$ with $k<1$. On the other hand, as shown from eqs. (\ref{eq:T2NEF}) and (\ref{eq:Lambda}), the antenna theorem forces the noise temperature of the receiver to scale with $1/f^2$ for a given field sensitivity. Therefore, the resulting noise temperature scales as $\mathrm{NET}\propto f^{2k - 2}$ which improves with frequency. 
	
	From Fig. \ref{fig:compa} we conclude that a free-space Rydberg electrometer with state-of-the-art sensitivity is not competitive with conventional receivers in terms of noise. However, the favorable trend with frequency reveals potential benefits at sub-millimeter wavelengths. Using optimally coupled cavities makes the atomic receivers competitive in terms of noise with conventional LNAs up to $\sim 30\,\mathrm{GHz}$. 

	
	$\quad$
	
	\subsubsection{Modeling optimal free-space sensitivity}
	
	We construct a Lindblad Master equation model of the system to examine the minimum possible free-space $\mathrm{NEF}_0$ of two-photon sensing using a microwave LO field. Results were generated using Rydberg sensing modeling software provided by the DEVCOM Army Research Laboratory and the Naval Air Warfare Center Weapons Division. Following the experimental setup described in \cite{cai2022}, we consider Rb85 in a $70\,\mathrm{mm}$ long vapor cell at $300\,\mathrm{K}$. We model a four-level system with an excitation scheme shown in the left of Fig. \ref{fig:simple}(b), using the excitation path $\ket{5S_{1/2,1/2}} \leftrightarrow \ket{5P_{3/2,1/2}} \leftrightarrow \ket{70S_{1/2,1/2}} \leftrightarrow \ket{70P_{3/2,1/2}}$. In addition to natural and thermal decays, $100\,\mathrm{kHz}$ of decoherence is added to each state, similar to the modeling in \cite{meyerPhysRevA.104.043103}. We assume the system is optically thin, and do not account for the effects of reduced optical intensities as the beams propagate through the media. We use $1\,\mathrm{mm}$ diameter beam waists, and calculate $\mathrm{NEF}_0$ assuming the photon-shot-noise of the probe beam is the only noise source. A full description of the $\mathrm{NEF}_0$ model calculation can be found in Appendix \ref{ap:ModLim}.
	
	Fig. \ref{fig:sencont} shows the calculated $\mathrm{NEF}_0$ for various probe and coupling laser Rabi frequencies $\Omega_P, \Omega_C$. There appear to be two local optima: one at $[\Omega_P,\Omega_C]\approx 2\pi\times[3, >5]$ MHz, and another with lower NEF$_0$ at $[\Omega_P,\Omega_C] \approx 2\pi\times[9.8, 1.8]$ MHz. At this $\Omega_P>\Omega_C$ optima the $\mathrm{NEF}_0 \approx0.5\uvmshz$, roughly a factor of two below the sensitivity achieved in \cite{cai2022}.
	
\begin{figure}[t]
	\centering
	\includegraphics[width = .45\textwidth]{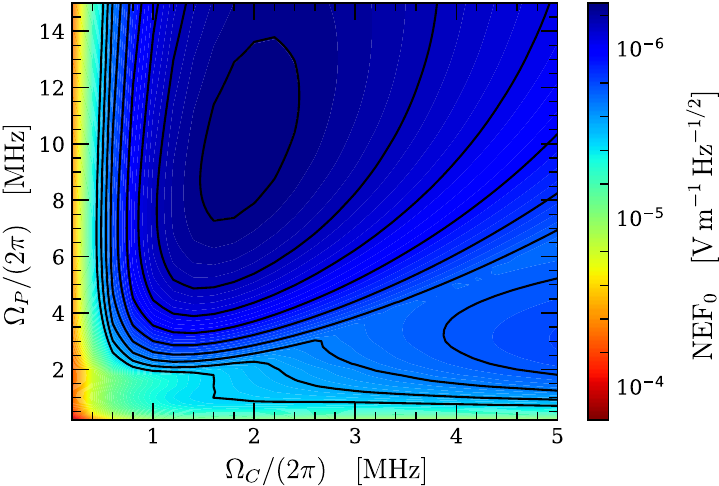}
	\caption{Modeled $\mathrm{NEF}_0$ for the $\ket{70P_{1/2,1/2}} \leftrightarrow \ket{70S_{3/2,1/2}}$ transition at various optical probe and coupling Rabi frequencies, assuming a $1\,\mathrm{mm}$ beam waist.}
	\label{fig:sencont}
\end{figure}
	
\begin{figure*}[t]
	\centering
	\includegraphics[width = .95\textwidth]{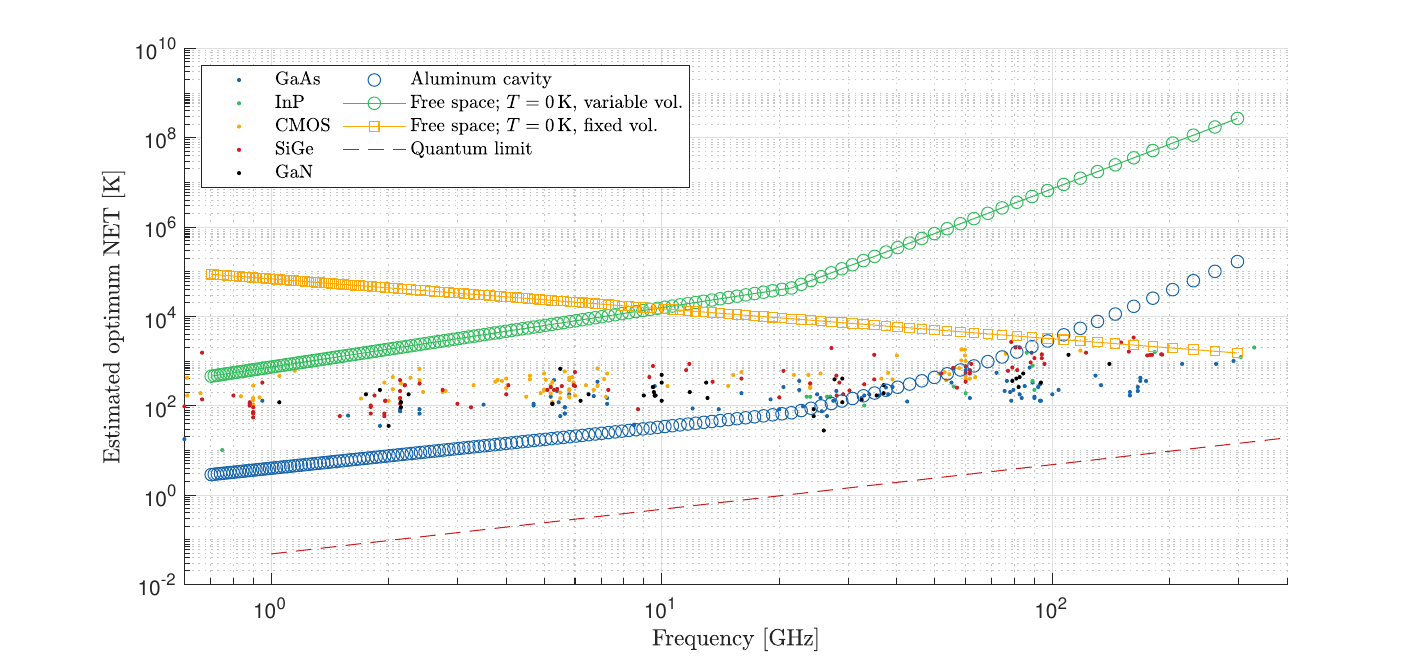}
	\caption{Noise temperature estimations of free-space and field-enhanced atomic sensors calculated by extrapolating the experimental results of \cite{cai2022} to other frequencies. Half-wavelength long rectangular waveguide resonators are used to enhance the field and the noise temperature is calculated via Eq. (\ref{eq:NEPwgmodel}) assuming $T_p=290\,\mathrm{K}$, $\Gamma_L=-1$ and $\sigma = 25\times 10^6\,\mathrm{S/m}$ bulk conductivity. The free-space noise temperature results are calculated using Eq. (\ref{eq:T2NEF}) assuming a dipole-like reception pattern ($G=3/2$) and neglecting the thermal background ($T=0$). Estimations in free-space where the interaction volume varies to match that of the equivalent waveguide resonator are also included. All estimations are compared with the noise temperature of LNAs reported in the literature \cite{belostotski}. }
	\label{fig:compa}
\end{figure*}

\begin{figure*}[t]
	\centering
	\includegraphics[width = .95\textwidth]{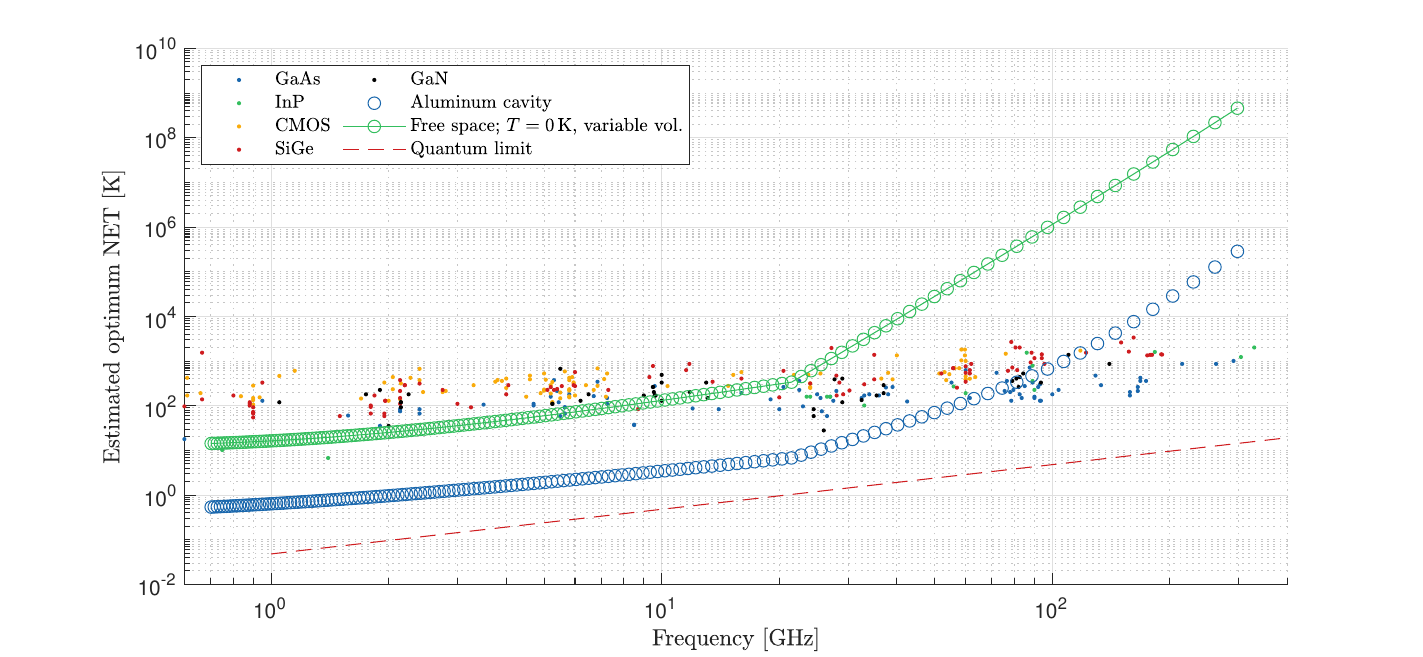}
	\caption{Same calculations done in Fig. \ref{fig:compa} but assuming optimal coupling and probe Rabi frequencies that minimize the $\mathrm{NEF}_0$ values.}
	\label{fig:compa10x}
\end{figure*}

	Fig. \ref{fig:compa10x} shows the results of extending this model to reproduce Fig. \ref{fig:compa} by adjusting the interaction lengths, beam widths, and target states to match target frequencies. We track the $\mathrm{NEF}_0$ of the $\Omega_P>\Omega_C$ local optimum shown in Fig. \ref{fig:sencont}. This regime has a lower $\mathrm{NEF}_0$ and is used in high-sensitivity laboratory demonstrations \cite{cai2022, jing2020atomic, prajapatirepump}. In contrast with prior analysis leading to Fig. \ref{fig:compa}, the model produces lower $\mathrm{NEF}_0$ values at $10.7\,\mathrm{GHz}$ and includes the effect of transit-time broadening as beam diameters shrink. The addition of transit-time broadening changes the approximate sensitivity dependence from $\mathrm{NEF}_0\propto 1/(w_0 L)$ to $\mathrm{NEF}_0\propto1/(w_0^2 L)$, though this dependence will be limited to regimes where transit-time broadening dominates over 
	other decoherence effects. 
	

\section{Discussion and conclusions}\label{sec:conclu}

	\begin{figure*}[ht]
		\centering
        \centering
        \begin{overpic}[width=0.95\textwidth,tics=10]{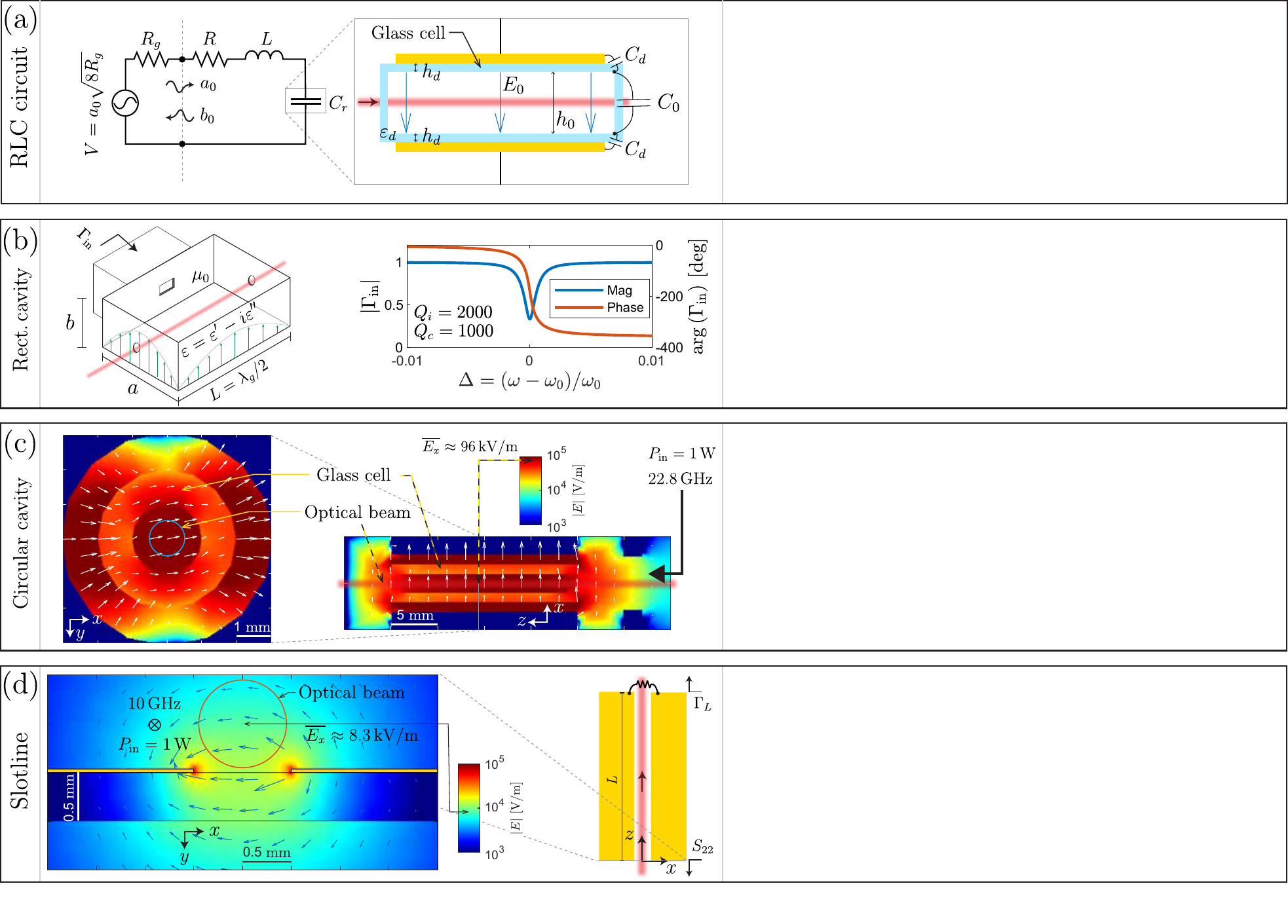}
\newcommand{\voffs}{0}
\newcommand{\voffall}{3}

        {\footnotesize 
       \newcounter{vloc}\setcounter{vloc}{66}
       \newcommand{\posleft}{51.5}
       \newcommand{\vsp}{-3}
        
        
        
        
        
        

        
        }
        
        {
        \small
        \setcounter{vloc}{66}
        \newcommand{\posleft}{57}
        \newcommand{\vsp}{-5}
\addtocounter{vloc}{\voffall}
        \put (76,\value{vloc}) {HO model:}\addtocounter{vloc}{\vsp}
        

        \put (\posleft,\value{vloc}) { $K = \frac{2\sqrt{2}}{\omega_0 h_0 C_0}\left(\frac{\sqrt{R_g}}{R + R_g}\right);\quad C = R_g/R$}\addtocounter{vloc}{\vsp}
        
        \put (\posleft,\value{vloc}) { ${R_g^\mathrm{(opt)}} = R\sqrt{1+\frac{8\Theta(f_0,T_p)}{\omega_0 ^2 R C_0^2 h_0^2 \mathrm{NEF}_0^2}}$}\addtocounter{vloc}{\vsp}

    

        }




        

        {
        \small
        \setcounter{vloc}{14}
        \newcommand{\posleft}{57}
        \newcommand{\vsp}{-4}
\addtocounter{vloc}{\voffall}
\addtocounter{vloc}{\voffs}
        \put (69,\value{vloc}) {WG model ($S_{22}=\Gamma_L=0$):}\addtocounter{vloc}{\vsp}

        \put (\posleft,\value{vloc}) { 
        $\frac{1}{B'}=e^{-2\alpha (L-2z)}
        ;\quad\frac{1}{C'}={2e^{2\alpha z}-e^{-2\alpha L}e^{4\alpha z}-1}$
        }\addtocounter{vloc}{\vsp}
        
        
        \put (\posleft,\value{vloc}) { 
        $\frac{1}{K'}=\frac{e^{\alpha z}}{K_W};\quad F\approx 3\sqrt{2G/3}$
        }\addtocounter{vloc}{\vsp}



        \put (\posleft,\value{vloc}) { 
        $\mathrm{NET}\approx 1\,\mathrm{K} + 2T_L + 1051\,\mathrm{K}\left(\mathrm{NEF_0/
        \frac{\mu V}{m\,\sqrt{Hz}}}\right)^2$
        }\addtocounter{vloc}{-2}
        

        }

        
        

        



        

        
        {
        \small 
        \setcounter{vloc}{49}
        \newcommand{\posleft}{57}
        \newcommand{\vsp}{-4}
\addtocounter{vloc}{\voffall}
\addtocounter{vloc}{\voffs}\addtocounter{vloc}{\voffs}
        \put (76,\value{vloc}) {HO model:}\addtocounter{vloc}{\vsp}\addtocounter{vloc}{-1}   

        \put (\posleft,\value{vloc}) { 
        $K^{2}=\frac{32\omega_{0}}{\pi ab\varepsilon'c^{2}}\left(\frac{Q_{i}Q_{c}}{Q_{i}+Q_{c}}\right)\Re^{-1}\left\{ \sqrt{\omega_{0}^{2}\mu_{0}\varepsilon-\left(\frac{\pi}{a}\right)^{2}}\right\} $
        }\addtocounter{vloc}{\vsp}

        \put (\posleft,\value{vloc}) { 
        $C=Q_i/Q_c$
        }\addtocounter{vloc}{\vsp}
        }



        
        

        
        {
        \small
        \setcounter{vloc}{33}
        \newcommand{\posleft}{57}
        \newcommand{\vsp}{-3}
\addtocounter{vloc}{\voffall}
\addtocounter{vloc}{\voffs}\addtocounter{vloc}{\voffs}
        \put (76,\value{vloc}) {HO model:}\addtocounter{vloc}{\vsp}\addtocounter{vloc}{-1}   

        \put (\posleft,\value{vloc}) { 
        $C\approx 1;\quad \quad Q_i \approx Q_c \approx 3170$
        }\addtocounter{vloc}{\vsp}\addtocounter{vloc}{-1}
        
        \put (\posleft,\value{vloc}) { 
        $K^2\approx 9.2\times 10^{9} \,\mathrm{V^2\,m^{-2}\,W^{-1}};\quad F\approx 16\sqrt{2G/3}$
        }\addtocounter{vloc}{\vsp}\addtocounter{vloc}{-1}
        
        \put (\posleft,\value{vloc}) { 
        $\mathrm{NET}\approx 2.19\,\mathrm{K} + 2T_p +7.86\,\mathrm{K}\left(\mathrm{NEF_0/
        \frac{\mu V}{m\,\sqrt{Hz}}}\right)^2$
        }\addtocounter{vloc}{-2}
        
        
        }

        \end{overpic}
		\caption{Application of the HO or WG models to various microwave structures for field confinement/enhancement: (a) A vapor cell inside a parallel plate capacitor on resonance with a series inductor forming an RLC circuit; (b) a half-wavelength rectangular cavity inserted inside a vapor cell; (c) circular waveguide sections with a cylindrical vapor cell embedded in a central near-cutoff region having uniform field and forming a half-wavelength long cavity; and (d) a matched slot waveguide on a quartz substrate inside a vapor cell. In all cases, the parameters needed to calculate the noise temperature of the receiver using either the HO or WG models are calculated analytically or from full-wave simulations. }
		\label{fig:table}
	\end{figure*}
	
In summary, we show that Rydberg atom-based electrometers embedded in port-coupled microwave structures can have better sensitivity than free-space configurations. In contrast to free-space coupled receivers, port-coupled receivers do not have the thermal-background lower bound, and in addition can enhance the electric field. 
Here we develop two theoretical approaches to quantify the noise temperature of port-coupled atomic receivers: (i) the harmonic oscillator (HO) model, suitable for arbitrary resonators where a single mode of relatively high quality factor is excited; and (ii) the waveguide (WG) model, suitable for non-resonant or resonant structures with arbitrary quality factors, as long as arbitrary guided modes are excited. The two models can be directly applied to practical cases, illustrated in Fig.\,\ref{fig:table}, discussed next.  

Electrically small field-enhancement resonators can be described as RLC circuits. In \cite{Hollowaysplitringdoi:10.1063/5.0088532}, two parallel plates confine the electric field around a vapor cell where Rydberg atoms are excited. This is equivalent to a capacitor $C_r$ that includes the glass walls and the atomic vapor region as shown in Fig.\,\ref{fig:table}(a). The circuit is brought to resonance at $\omega_0$ with an inductance $L$ and dissipation losses are accounted for through the resistance $R$. The circuit is driven by a voltage generator $V=a_0\sqrt{8R_g}$ with internal resistance $R_g$ and available power $\abs{a_0}^2$. The power delivered to the resonator (and thus, the magnitude of the electric field sensed by the atoms $E_0$) is maximized when $R_g=R$. In \cite{Hollowaysplitringdoi:10.1063/5.0088532}, the inductance $L$ of a sub-wavelength (non-resonant) loop antenna is leveraged to form the series RLC circuit shown in Fig \ref{fig:table}(a). Dissipation losses are included in $R$, while $R_g$ becomes the radiation resistance of the loop. Interestingly, $E_0$ is independent on the separation between plates given by the cell height and glass wall thickness, $h = h_0+2h_d$. The reason is that, while $E_0\approx V/h$ for a given voltage $V$ across the capacitor, the capacitance $C_r\propto 1/h$ and thus $V\propto h$ at resonance, since the current is fixed by the resistor values. If a non-reactive antenna with a parallel inductor were used instead, the resulting parallel RLC resonant circuit  would have $E_0 \propto 1/h$. 

For the circuit in Fig.\,\ref{fig:table}(a), the intrinsic and coupling quality factors are given by \cite{pozar2011microwave} $Q_i=\left(\omega_0 R C_r\right)^{-1}$ and $Q_c=\left(\omega_0 R_g C_r\right)^{-1}$ respectively, where $\omega_0=\left(LC_r\right)^{-1/2}$ is the resonance frequency. Since the energy stored by a resonator decays exponentially with a rate $\omega_0/Q_{\ell}$, where $Q_\ell$ is the loaded quality factor, the field coupling and loss rates $\gamma_c$ and $\gamma_i$, respectively, are given by 

\begin{equation}
    \gamma_{c,i}=\frac{\omega_0}{2Q_{c,i}}. 
\end{equation}
Since the RLC resonator is zero-dimensional, the free spectral range is infinite leading to $\tau_g\rightarrow\tau\rightarrow 0$. Therefore, we can set $\tau_g/\tau = 1$ in Eq. (\ref{eq:Ksq}) which can be also proven by directly calculating the electric field squared magnitude per unit stored energy. 
%
The righthand column of Fig.\,\ref{fig:table}(a) show the parameters needed to calculate the noise temperature of the field-enhanced receiver  in terms of the circuit elements, using the HO model. This is done via Eq. (\ref{eq:NEP}) after setting $T_A=0$ and dividing by $k_B$.

To illustrate the usefulness of the HO model to analyze high-Q arbitrary resonators, consider the rectangular half-wavelength cavity shown in Fig. \ref{fig:table}(b). If coupling is performed via a small perturbation such as e.g. an iris, the mode profile is almost undisturbed and the squared field amplitude per unit stored energy $K_U^2$, as well as the phase and group velocity ratio $\tau_g/\tau$ can be found analytically. Full-wave solvers, such as Ansys HFSS, can also be used for more accuracy. The intrinsic (unloaded) $Q_i$ and coupling (external) $Q_c$ quality factors can be found numerically using full-wave simulations. One way this can be done is by directly obtaining the loaded quality factor $Q_\ell$ of the eigensolutions of the coupled cavity. Next, $Q_c$ is obtained by repeating the simulation after eliminating any losses. Then, $Q_i$ is found as $Q_i=\left(Q_\ell^{-1} - Q_c^{-1}\right)^{-1}$. Another way to obtain $Q_i$ and $Q_c$ is using a driven full-wave simulation. This is done by fitting magnitude and phase of the frequency-dependent port reflection coefficient $\Gamma_\mathrm{in}$ to the theoretical solution $b_1$ of Eq. (\ref{eq:caveq1}) in the high-Q limit:

\begin{equation}
    \Gamma_{\mathrm{in}}=\frac{({1-C})/({1+C})+i2Q_{\ell}{\Delta}}{1+i2Q_{\ell}\Delta},
\end{equation}
where $\Delta = (\omega-\omega_0)/\omega_0$, $C = Q_i/Q_c$ and $Q_\ell = {Q_i Q_c}/{(Q_i+Q_c)}$. An example of the frequency response of $\Gamma_\mathrm{in}$ for $Q_i=2000$ and $Q_c=1000$ is shown in Fig. \ref{fig:table}(b).  Using the analytical solutions for the $\mathrm{TE}_{101}$ mode in a rectangular cavity, we obtain 

\begin{equation}
    \frac{\tau_g}{\tau} = \frac{\omega_0^2}{c^2\Re^2\left\{\sqrt{\omega_0^2\mu_0\varepsilon-\left(\frac{\pi}{a}\right)^2}\right\}},
\end{equation}
and, using Eq. (\ref{eq:Ku})

\begin{equation}
    K_U^2 = \frac{8\Re\left\{\sqrt{\omega_0^2\mu_0\varepsilon-\left(\frac{\pi}{a}\right)^2}\right\}}{\varepsilon'\pi ab}.
\end{equation}
These parameters are then used to find the coefficient $K$ (Eq. (\ref{eq:Ksq})) which, together with $C$, is used to calculate the noise temperature of the receiver via Eq. (\ref{eq:NEP}). This is shown in the rightmost column of Fig. \ref{fig:table}(b).

The same method described above to obtain $Q_i$ and $Q_c$ can be used in 
%
%
non-canonical geometries such as the cavity shown in Fig. \ref{fig:table}(c).  
In this case, the cavity is similar to a half-wavelength long circular waveguide resonating at $\omega_0 = 2\pi\times 22.8\,\mathrm{GHz}$. A $4\,\mathrm{mm}$ diameter, $1\,\mathrm{mm}$ wall thickness, and $20\,\mathrm{mm}$ long cylindrical fused silica vapor cell is inserted inside a near-cutoff section of circular waveguide. The resulting long guided wavelength provides a good electric field uniformity ($\mathrm{VSWR}\approx 1.5$) along the vapor cell length, where the atoms are illuminated with the laser beams. This electrically short waveguide is then terminated by near quarter wavelength sections, one of which has an aperture for microwave coupling.  
From the simulated on-resonance field distribution within the vapor cell with an incident microwave power $P_\mathrm{in}=1\,\mathrm{W}$, the parameter $K$ of Eq. (\ref{eq:Ksq}) is readily available. This corresponds to a field enhancement factor $F\approx 16$ when the input port is connected to an electrically-small antenna with gain $G=3/2$. In this particular example, the cavity is critically coupled, so $C=1$, and the intrinsic quality factor is $Q_i\approx 3170$. The resulting $\mathrm{NET}$ is found from Eq. (\ref{eq:NEP}) after setting $T_A=0$ and dividing by $k_B$, leading to the convenient expression in the rightmost column of Fig. \ref{fig:table}(c).  The strong field enhancement factor significantly mitigates the effect of the noise floor $\mathrm{NEF}_0$ that the atoms would exhibit in free-space. For $\mathrm{NEF}_0$ values on the order of a few $\uvmshz$,  the receiver noise is dominated by the thermal contribution of the resonator at room temperature $T_p$. In these cases, the optimal noise performance is achieved by overcoupling the cavity $C = Q_i/Q_c>1$ to a value given by Eq. (\ref{eq:gammaopt}). Since $K^2$ was obtained from a critically-coupled simulation at resonance, from Eq. (\ref{eq:Ksq}) and the fact that $2\gamma_i=\omega_0/Q_i$,

\begin{equation}
    \frac{\tau_g}{\tau}K_U^2=\frac{\omega_0}{Q_i}K^2\approx 4.16\times 10 ^ {17}\,\mathrm{V^2\,m^{-2}\,J^{-1}},
\end{equation}
which is independent from $Q_c$. Therefore, from the simulated field distribution at a single coupling condition, the values of $K^2$ at any other coupling strength can be calculated via Eq. (\ref{eq:Ksq}), and the optimal coupling via Eq. (\ref{eq:gammaopt}). For this particular cavity, the minimum noise temperature is $\mathrm{NET}\approx 14.7\,\mathrm{K}$ for $\mathrm{NEF}_0=0.2\uvmshz$, $\mathrm{NET}\approx 72.8\,\mathrm{K}$ for $\mathrm{NEF}_0=1\uvmshz$, and $\mathrm{NET}\approx 1187\,\mathrm{K}$ for $\mathrm{NEF}_0=10\uvmshz$.

Non-resonant but confining guiding structures have also been proposed in e.g. \cite{MeyerPhysRevApplied.15.014053} to improve the responsivity of the atoms in a port-fed configuration. In Fig. \ref{fig:table}(d) the fields of a planar slotted line waveguide on a quartz substrate at $10\,\mathrm{GHz}$ are computed with a full-wave solver (HFSS), and normalized to a $P_\mathrm{in}=1\,\mathrm{W}$ excitation. The complex propagation constant $k_z =\alpha+i\beta$ is also obtained from the simulations, resulting in

\begin{equation}
    \kappa_z = (0.08+i241.2)\,\mathrm{m^{-1}}.
\end{equation}

The WG model is suitable in this case, and the parameter $K_W$ in Eq. (\ref{eq:KW}) is readily available from the cross-section field distribution of the propagating mode shown in Fig. \ref{fig:table}(d) for $P_\mathrm{in}=1\,\mathrm{W}$ input power. In this case, the $x$-component of the electric field averaged across the Gaussian distribution of a $1\,\mathrm{mm}$ diameter laser beam $\overline{E_x}$ is used, leading to

\begin{equation}
    K_W = \frac{\overline{E_x}}{P_\mathrm{in}}\approx 8.3\times 10^3\,\mathrm{V\,m^{-1}\,W^{-1/2}}. 
\end{equation}

 The resulting expressions for the coefficients $B'$, $K'$, and $C'$, shown in the rightmost column of Fig. \ref{fig:table}(d) are heavily simplified because the waveguide of length $L$ is terminated with a matched load, and is fed by a matched port ($S_{22}=\Gamma_L=0$). When $\alpha L \ll 1$, these coefficients are approximately constant in space, leading to the handy $\mathrm{NET}$ expression shown in the figure. This is the input-referred noise temperature of the receiver and is calculated via Eq. (\ref{eq:NEPwgmodel}) after setting $T_A=0$ and dividing by $k_B$. The main contributors to the noise of the system are the physical temperature $T_L$ of the matched load, and the field noise floor of the atoms in free-space $\mathrm{NEF}_0$. The latter adds about $1051\,\mathrm{K}$ when $\mathrm{NEF}_0 = 1\uvmshz$, which is significant. This is due to the low field enhancement factor $F\approx 3\sqrt{2G/3}$ of the slotted-line waveguide when fed with an antenna of gain $G$. 

In summary, Fig.\,\ref{fig:table} is a stencil for designing field-enhancement microwave structures that minimize noise of Rydberg-atom receivers. The results in this figure are obtained from the detailed derivations given in the paper and appendices. Overall conclusions from this study are summarized as follows.

\begin{itemize}
   \item The noise floor limits of Rydberg electrometers in a free-space configuration are quantified, taking into account thermal black-body radiation and vacuum fluctuations at microwave frequencies, for both microwave homodyne and heterodyne LO receivers. It is found that these limits are just a few times below the state-of-the-art sensitivities demonstrated experimentally to date.

   
   \item It was found that instead of free-space atomic receiver configurations, port-coupled microwave structures that confine or enhance the field sensed by the atoms can be used to reach receiver temperature levels below the $290\,\mathrm{K}$ room temperature.
    
    
   \item Two theoretical models are developed to calculate the noise temperature of atomic receivers enhanced by resonant and/or confining microwave structures. The models are general and can be applied to arbitrary structures, whose field distributions are extracted from full-wave simulations. Microwave thermal and quantum noise are accounted for, establishing lower bounds to the achievable noise floor of the receivers. 
   
    
    \item We compare the electric field sensitivity of free-space coupled atomic receivers with the noise temperature of conventional microwave frontends in a consistent manner. Even optimistically, current sensitivity levels demonstrated experimentally in X-band are far from being competitive with what could be achieved with an off-the-shelf LNA. However, by confining the fields in simple metallic rectangular waveguide resonators, the resulting noise temperature of the atomic receiver becomes competitive with existing LNAs.
    
    \item The noise temperature of  
    existing LNAs at frequencies ranging from $600\,\mathrm{MHz}$ to $330\,\mathrm{GHz}$, is compared with the equivalent noise temperature of atomic receivers enhanced by metallic rectangular waveguide resonators. To this end, the state-of-the-art sensitivity value demonstrated in X-band is extrapolated to other microwave frequencies, taking into account the changes in interaction volume and transition dipole moment. It is found that, by simply using rectangular cavities to enhance the field, current atomic receivers become competitive with existing LNAs up to about K-band. Furthermore, using optimal coupling and probe beam Rabi frequencies would make cavity-enhanced atomic receivers competitive up to W-band. 
    \item Interestingly, the noise temperature of free-space configurations with a constant interaction volume improves as frequency increases. If black body radiation noise could be avoided by e.g., directing the received RF fields towards a cold source using reflectors, free-space configurations with current sensitivities would be competitive with LNAs beyond W-band. 
    
\end{itemize}


\section{Acknowledgments}\label{sec:ack}
{This work was funded by Lockheed Martin Corporation, under award number MRA17-003-RPP028, and ColdQuanta through the DARPA SaVANT program, under award number 7051-SC-CU-P1. The authors acknowledge Drs.\,Y.J.\,Wang, H.\,Fan and S.\,Roof of ColdQuanta for helpful technical discussions, Prof. Josh Combes from CU Boulder for valuable comments, as well as Dr.\,S.\,Subramanian from Lockheed Martin and Dr.\,T.\,\'Cur\v ci\'c from DARPA for encouraging this work.}

		\appendix
		
\section{Reception pattern and effective aperture of free-space-coupled atomic receivers}\label{ap:F}	

\begin{figure}[t]
	\centering
	\includegraphics[width = 0.4\textwidth]{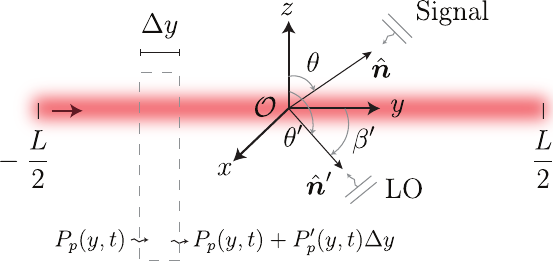}
	\caption{Two plane waves (LO and signal) interfere within the interaction volume of a Rydberg electrometer. 
	}
	\label{fig:gainscheme}
\end{figure}

Consider the probe beam propagates along the $y$-axis and interrogates the atoms within the region $-L/2\leq y\leq L/2$ as shown in Fig. \ref{fig:gainscheme}. Suppose the diameter of the beam is electrically small and the atoms are responsive to the $z$-component of the electric field. Signal and local oscillator plane waves propagate in the directions $-\hat{\boldsymbol{n}}(\theta,\phi)$ and $-\hat{\boldsymbol{n}}'(\theta',\phi')$ and have electric fields $\boldsymbol{E}$ and $\boldsymbol{E}'$ respectively. The complex amplitude of the signal is a slow-varying process and is assumed constant within the interaction volume, which is valid when the volume is much smaller than the coherence length  $L\ll c/\Delta f$ determined by the post-detection bandwidth $\Delta f$. The total electric field in every point in space $\boldsymbol{r}$ is given by

\begin{equation}
    \boldsymbol{E}_{t}=\frac{1}{2}\boldsymbol{E}e^{i\kappa\hat{\boldsymbol{n}}\cdot\boldsymbol{r}}e^{i\omega_{\mathrm{RF}}t}+\frac{1}{2}\boldsymbol{E}'e^{i\kappa\hat{\boldsymbol{n}}'\cdot\boldsymbol{r}}e^{i\omega_{\mathrm{LO}}t}+\mathrm{c.c.},
\end{equation}
where $\kappa=2\pi/\lambda_0$,  $\boldsymbol{E}=E_{\theta}(t)\hat{\boldsymbol{a}}_{\theta}+E_{\phi}(t)\hat{\boldsymbol{a}}_{\phi}$ and $\boldsymbol{E}'=E_{\theta}'(t)\hat{\boldsymbol{a}}_{\theta}'+E_{\phi}'(t)\hat{\boldsymbol{a}}_{\phi}'$. The atoms are responsive to the IF beatnote $E_b(\boldsymbol{r},t)$ generated by $\boldsymbol{E}_t\cdot \hat{\boldsymbol{a}}_z$, i.e., 

\begin{equation}
\begin{split}
E_{b}(\boldsymbol{r},t)	&\propto\left(\hat{\boldsymbol{a}}_{z}\cdot\boldsymbol{E}e^{i\kappa\hat{\boldsymbol{n}}\cdot\boldsymbol{r}}e^{i\omega_{\mathrm{RF}}t}\right) \\ &\quad\times\left(\hat{\boldsymbol{a}}_{z}\cdot\boldsymbol{E}'^{*}e^{-i\kappa\hat{\boldsymbol{n}}'\cdot\boldsymbol{r}}e^{-i\omega_{\mathrm{LO}}t}\right)+\mathrm{c.c.} \\
	&\propto E_{\theta}'^{*}\sin\theta'E_{\theta}\sin\theta e^{i\kappa(\hat{\boldsymbol{n}}-\hat{\boldsymbol{n}}')\cdot\boldsymbol{r}}e^{i\omega_{\mathrm{IF}}t}+\mathrm{c.c.}
\end{split}\label{eq:beatnoter}
\end{equation}
Along the interaction region $\boldsymbol{r}=y\hat{\boldsymbol{a}}_y$, and for a fixed LO amplitude $E_\theta '$ and direction $\theta'$, the beatnote of Eq. (\ref{eq:beatnoter}) takes the following form after putting $E_\theta =\left|E_{\theta}\right|e^{i\varphi(t)}$:

\begin{multline}
   E_{b}(y,t)\propto\sin\theta e^{i\kappa(\sin\theta\sin\phi-\sin\theta'\sin\phi')y} \\ \times \left|E_{\theta}\right|e^{i\varphi(t)}e^{i\omega_{\mathrm{IF}}t}+\mathrm{c.c.}
\end{multline}

\begin{figure}[t]
	\centering
	\includegraphics[width = 0.49\textwidth]{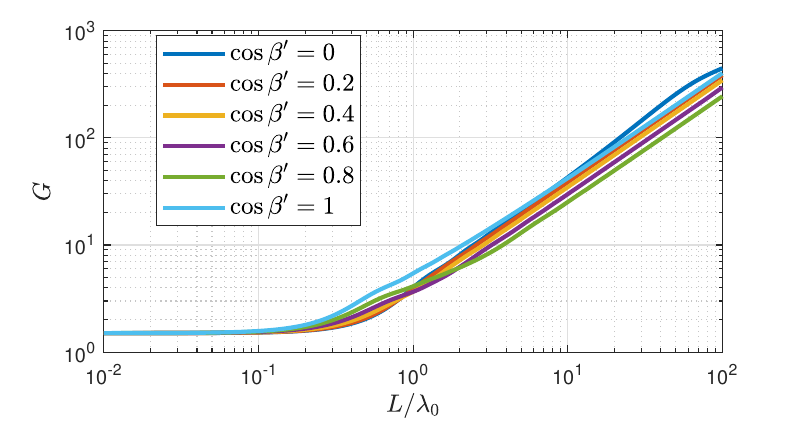}
	\caption{Gain of a free-space coupled atomic ensemble as a function of the wavelength-normalized interaction length. Curves are plotted for different LO incidence angles $\beta'$ with respect to the axis of the optical beam. 
	}
	\label{fig:gains}
\end{figure}

The atoms at $y$ respond to the local microwave beatnote field $E_{b}(y,t)$ by attenuating the transmitted probe power $P_p(y+\Delta y, t)\approx P_p(y,t)+\dv{}{y}P_p(y,t)\Delta y$ by a factor proportional to 

\begin{equation}
    \dv{}{y}P_p(y,t) \propto E_{b}(y,t).
\end{equation}
Therefore, in the small-signal regime, the transmitted probe at the end of the interaction has a beatnote-modulated (AC) component that can be approximated by

\begin{multline}
    P_{\mathrm{sig}}(t)\propto\left|E_{\theta}\right|e^{i\varphi(t)}e^{i\omega_{\mathrm{IF}}t} \\ \times \sin\theta\int_{-\frac{L}{2}}^{\frac{L}{2}}e^{i\kappa(\sin\theta\sin\phi-\sin\theta'\sin\phi')y}\,\mathrm{d}y +\mathrm{c.c.}
\end{multline}
Solving the integral, we obtain

\begin{multline}
    P_{\mathrm{sig}}(t)\propto \left|E_{\theta}(t)\right|\cos\left[\omega_{\mathrm{IF}}t+\varphi(t)\right] \\ \times \underbrace{ \sin\theta\mathrm{sinc}\left[\frac{\kappa L}{2}(\sin\theta\sin\phi-\sin\theta'\sin\phi')\right]}_{\mathsf{F}(\theta,\phi)},
\end{multline}
where $\mathrm{sinc}(x)=\sin(x)/x$. $\mathsf{F}(\theta,\phi)$ is the ``reception pattern'' of the atomic receiver. From this, the effective reception gain $G$ of the atomic ensemble can be calculated as $G=4\pi/\int_0^{2\pi}\int_0^\pi \mathsf{F}^2(\theta,\phi)\sin\theta\,\mathrm{d}\theta\mathrm{d}\phi$, in terms of the electrical length $L/\lambda_0=\kappa L/(2\pi)$ for a given LO incidence angle $(\theta',\phi')$. Note that $\cos\beta'=\sin\theta'\sin\phi'$ is the direction cosine of the LO angle of arrival with respect to the axis of the beam. These calculations of $G$ are plotted in Fig. \ref{fig:gains}. Note how $G\rightarrow 3/2$ as $L/\lambda_0\rightarrow 0$.

Note that from the calculations above, for a fixed LO angle of arrival, the complex amplitude of the photodetected beatnote that modulates the probe is proportional to the electric field complex amplitude of the incoming signal plane wave. This linear relationship is identical to that of the complex amplitude seen at the port of an antenna, and thus, the atomic receiver is mathematically indistinguishable from a receiving antenna with radiation pattern $\mathsf{F}(\theta,\phi)$. We can then expect the antenna theorem \cite{oliver} to be applicable to the atomic receiver, and the ``effective area'' that intercepts the power from an incoming plane wave with angle of arrival $(\theta,\phi)$ to be $A_\mathrm{eff}=\frac{\lambda_0^2}{4\pi}G\mathsf{F}^2(\theta,\phi)$. This implies that, if the receiver is in a thermal background at temperature $T_A$, the noise floor is fixed to $\Theta(f,T_A)$ while the signal originating from an incoming plane wave in the direction where $\mathsf{F}$ is maximum, is proportional to $G$. Since the results of Section \ref{sec:free} were derived in the electrically-small interaction volume limit (where $G=3/2$), for an arbitrary volume leading to $G> 3/2$, the $\mathrm{NEF}_\mathrm{ex}$ values can still be obtained from Fig. \ref{fig:nefplot} after dividing by $\sqrt{2G/3}$.

\section{Calculation of stored energies}\label{ap:A}

The intra-cavity stored energy can be calculated by driving Eq. (\ref{eq:rateeq}) with a monochromatic excitation until steady-state is reached, then switching off the input and integrating the reflected power via Eq. (\ref{eq:caveqb1}) plus the dissipated power $(1-\left|{\satt}\right|^2)\left|b_2\right|^2$. The result of this thought experiment is that the stored energy equals $\tau_g \left|b_2\right|^2$, which is intuitive as the intra-cavity power circulates during the interval $\tau_g$. 

Suppose one desires to calculate the intra-cavity energy spectral density due to an input PSD $\abs{a_1(\omega_0+\delta\omega)}^2$. 
To sample the spectral response of the resonator's energy at a given detuning $\delta\omega$, the observation bandwidth can be made arbitrarily narrow such that Eq. (\ref{eq:rateeq}) is solved in steady state ($b_2'(t)=0$). The resulting signal-related \emph{energy spectral density} $\expval{w_\mathrm{sig}(\delta\omega)}=\tau_g\abs{b_2(\delta\omega)}^2$ approximates a Lorentzian shape $H_L(\omega)$ for $\delta\omega \ll \tau^{-1}$ 

\begin{equation}
	\begin{split}
		\expval{w_\mathrm{sig}(\delta\omega))}
		&=\frac{\tau_g\left(1-e^{-2\gamma_c\tau}\right)}{\left(1-e^{-\gamma\tau}\right)^2+4e^{-\gamma\tau}\sin^2\left(\frac{\delta\omega\tau}{2}\right)}\abs{a_1}^2 \\
		&\approx \frac{\tau_g\left(1-e^{-2\gamma_{c}\tau}\right)}{\left(1-e^{-\gamma\tau}\right)^{2}}\underbrace{\left[1+\left(\pi\frac{\delta\omega}{\Delta\omega_{r}}\right)^{2}\right]^{-1}}_{H_{L}(\omega_{0}+\delta\omega)}\abs{a_1}^2
	\end{split}
\end{equation}
where $\Delta\omega_r= 2\pi\Delta f_r =\int_{-\infty}^{\infty}H_L(\omega)\,\mathrm{d}\omega$ is the power equivalent bandwidth of the resonance, and $\Delta f_r=\sinh(\gamma\tau/2)/\tau=\frac{\pi}{2}B_\mathrm{3dB}$, where 
$B_\mathrm{3dB}$ 
is the FWHM linewidth.  

Let us now calculate the stored energy spectral density due to thermal excitation. Two contributions are to be considered: the thermal background at temperature $T_A$ observed by the antenna, and the thermal noise generated by the lossy resonator at physical temperature $T_p$. The former produces an input $a_1=n_a$ having a PSD\footnote{We include vacuum fluctuations (arising from signal and image bands in case of microwave heterodyning) by using Eq. (\ref{eq:CallenWelton}). }  $\abs{n_a(\delta\omega)}^2 = \Theta(f,T_A)$, and the latter can be modeled as a Langevin (primary) white noise source $n_p(t)$ added to $a_2(t)$ such that $a_2(t) = \satt\exp(-i\omega\tau)b_2(t-\tau_g) + n_p(t)$. The PSD of the Langevin source is \footnote{Resonator loss couples vacuum fluctuations and thermal noise in the same way.}  $\abs{n_p(\delta\omega)}^2 = K_p\Theta(f,T_p)$ where $K_p$ is a proportionality constant that is independent of temperatures and we shall determine. Both thermal sources $a_p$ and $a_1$ are uncorrelated and their incorporation transforms Eq. (\ref{eq:rateeq}) into

\begin{multline}
	b_2'(t)  +	\frac{e^{(\gamma + i\delta\omega)\tau} - 1}{\tau_g}{b_2(t)}{}  =     \frac{{\Sko}}{\tau_g}e^{(\gamma + i\delta\omega)\tau}n_a(t) \\+ \frac{{\Srt}}{\tau_g}e^{(\gamma + i\delta\omega)\tau}n_p(t).
	\label{eq:rateeqLang}
\end{multline}
Again, the power spectral density $\abs{b_2(\delta\omega)}^2$ is sampled by solving Eq. (\ref{eq:rateeqLang}) in steady state. Since $n_a(t)$ and $n_p(t)$ are uncorrelated, the thermal-related stored energy spectral density $\expval{w_\mathrm{th}(\delta\omega)}=\tau_g\abs{b_2(\delta\omega)}^2$ is computed using superposition of contributing PSDs resulting

\begin{multline}
\expval{w_\mathrm{th}(\delta\omega)} =\frac{\tau_g\left(1-e^{-2\gamma_{c}\tau}\right)}{\left(1-e^{-\gamma\tau}\right)^{2}}H_{L}(\omega_{0}+\delta\omega)\Theta(f,T_A) \\ + \frac{\tau_g e^{-2\gamma_{c}\tau}}{\left(1-e^{-\gamma\tau}\right)^{2}}H_{L}(\omega_{0}+\delta\omega)K_{p}\Theta(f,T_{p}). \label{eq:ESDthermal}
\end{multline}

To find the value of $K_p$, suppose $T_p=T_A=T$. Since the system is in thermal equilibrium, the total energy stored in the cavity in the mode of interest must be $\Theta(f, T)$ \cite{oliver}. After integrating both sides of Eq. (\ref{eq:ESDthermal}), the result $\int_{-\infty}^\infty\expval{w_\mathrm{th}(\delta\omega)}\,\mathrm{d}\delta\omega=\Theta(f, T)$ iif 

\begin{equation}
	K_p=e^{2\gamma_c\tau}\left[2\left(1-e^{-\gamma\tau}\right)e^{-\gamma\tau/2}-\left(1-e^{-2\gamma_c\tau}\right)\right].
\end{equation}

The last step is to associate the free-space $\mathrm{NEF}_0$ to a given stored energy spectral density. Suppose the field distribution of the eigenmode of interest is $\boldsymbol{\Psi}(\boldsymbol{r})$. The scalar function $K_{U}(\boldsymbol{r}_{0})$ given by Eq. (\ref{eq:Ku}) 
is the magnitude of the electric field at $\boldsymbol{r}_{0}$ of the eigenmode normalized to unity energy in the direction of observation $\hat{\boldsymbol{a}}_{e}$. Therefore, a mode 
whose squared-field spectral density at $\boldsymbol{r}_{0}$ is $\mathrm{NEF}_0^2$, will correspond with an energy spectral density $\expval{w_0(\delta\omega)}$ given by 

\begin{equation}
	\expval{w_0(\delta\omega)} = \frac{\mathrm{NEF}_0^2}{K_U^2(\boldsymbol{r}_{0})},
	\label{eq:w0ESD}
\end{equation}
which is independent from $\delta\omega$ because other modes are not excited within a small detuning $\delta\omega \ll \tau^{-1}$. Equation (\ref{eq:w0ESD}) is the noise floor of the thermally unpopulated atomic receiver in free space ($\mathrm{NEF}_0^2$) referred to stored energy spectral density. This is thus directly comparable with the thermally induced noise floor of Eq. (\ref{eq:ESDthermal}). After setting the signal-to-noise ratio to unity, the Noise Equivalent Power (NEP) results

\begin{equation}
\mathrm{NEP}	=\Theta(f,T_{A})+\frac{1}{C}\Theta(f,T_{p})+\frac{1}{K^2}\mathrm{NEF}_{0}^{2},
\end{equation}
where

\begin{equation}
C^{-1}=\frac{2\left(1-e^{-\gamma\tau}\right)e^{-\gamma\tau/2}}{1-e^{-2\gamma_{c}\tau}}-1\approx\frac{\gamma_{i}}{\gamma_{c}},
\end{equation}
and

\begin{equation}
	\begin{split}
	K^{2}&=\frac{\tau_g\left(1-e^{-2\gamma_{c}\tau}\right)}{\left(1-e^{-\gamma\tau}\right)^{2}}K_{U}^{2}(\boldsymbol{r}_{0})H_{L}(\omega_{0}+\delta\omega) \\ &\approx\frac{2\gamma_{c}}{\left(\gamma_{i}+\gamma_{c}\right)^{2}}\frac{\tau_g}{\tau}K_{U}^{2}(\boldsymbol{r}_{0})H_{L}(\omega_{0}+\delta\omega),
	\end{split}
\end{equation}
	where the approximations hold as $\gamma_{i,c}\ll\tau^{-1}$.

	\section{Spatial dependence of PSD due to differential noise source}\label{ap:B}

The equations governing the \emph{total} waves that are a consequence of the $ \diplus $ emanating from $\Delta V$ are, according to Fig. \ref{fig:JiMi},

\begin{equation}
	\begin{split}
		v_2 &= u_2\Gamma_{L} e^{-2\kappa_z(L-z')} \\
		u_2 &= u_1+ \diplus  \\
		u_1 &= v_1 \Srt e^{-2\kappa_z z'} \\
		v_1 &= v_2,
	\end{split}\label{eq:uv}
\end{equation}
where the attenuation $e^{-2\alpha\Delta z}$ within $\Delta V$ has been neglected as including it would generate higher order terms of $\Delta z$ that vanish as $\Delta z\rightarrow 0$ which will be done later. The solutions to Eq. (\ref{eq:uv}) are

\begin{equation}
\begin{split}
	u_2 &= \frac{1}{1-{\Srt}\Gamma_{L}e^{-2\kappa_z L}} \diplus  \\
	u_1 &= \left(\frac{1}{1-{\Srt}\Gamma_{L}e^{-2\kappa_z L}}-1\right) \diplus  \\
	v_2 &= v_1 = \frac{\Gamma_{L} e^{-2\kappa_z(L-z')}}{1-{\Srt}\Gamma_{L}e^{-2\kappa_z L}} \diplus . 
\end{split}\label{eq:uvsol}
\end{equation}

For $z>z'$ the total field distribution reads $A^+(z,z') = u_2e^{-\kappa_z(z-z')}+ v_2e^{\kappa_z(z-z')}$,  whereas for $z<z'$, $A^+(z,z') = u_1e^{-\kappa_z (z-z')}+ v_1e^{\kappa_z (z-z')}$. Using the solutions of Eq. (\ref{eq:uvsol}), we obtain $A^+(z,z')=d_+^if^+(z,z')$ and $A^-(z,z')=d_-^if^-(z,z')$, where

\begin{multline}
f^{+}(z,z')=\frac{1}{1-{\Srt}\Gamma_{L}e^{-2\kappa_{z}L}}\\
\times\begin{cases}
	e^{\kappa_{z}z'}\left[e^{-\kappa_{z}z}+\Gamma_{L}e^{-2\kappa_{z}L}e^{\kappa_{z}z}\right] & z>z'\\
	\Gamma_{L}e^{-2\kappa_{z}L}e^{\kappa_{z}z'}\left[e^{\kappa_{z}z}+{\Srt}e^{-\kappa_{z}z}\right] & z<z'
\end{cases},\label{eq:fplus}
\end{multline}
and $f^{-}(z,z')$ is obtained from $f^{+}(z,z')$ after substituting $z\rightarrow-z$, $z'\rightarrow-z'$, $\Gamma_{L}e^{-2\kappa_{z}L}\rightarrow {\Srt}$, and ${\Srt}\rightarrow\Gamma_{L}e^{-2\kappa_{z}L}$.
$A^\pm(z,z')$ corresponds to the total field observed at $z$, due to the differential thermal contribution at $z'$ emanating in the direction of $ \diplusminus $. The derivations shown here use traveling wave amplitudes which are defined in terms of the transversal components of the electric field of the waveguide modes. If the atoms in a given setup  are sensitive to the longitudinal components of the electric field, then the substitutions $\Gamma_L\rightarrow -\Gamma_L$ and ${\Srt}\rightarrow -{\Srt}$ must be performed in eqs. (\ref{eq:uv}), (\ref{eq:uvsol}) and (\ref{eq:fplus}). 


%
%

Since $A^+$ and $A^-$ are uncorrelated, their individual powers can be added to find the total differential PSD distribution $\mathrm{d}{\Sp}(z,z')$, given by Eq. (\ref{eq:dSpzzp}).

\section{Dependence of free-space sensitivity on interaction volume}\label{ap:C}	
Experimental data suggests that the state-of-the-art field sensitivity values demonstrated so far \cite{prajapatirepump, cai2022} are limited by the photon shot noise of the optical detection scheme. In this case, the sensitivity of the atomic receiver scales as 

\begin{equation}
    \mathrm{NEF_0}\propto \left(m_p\sqrt{P_0}\right)^{-1},\label{eq:NEFtrend}
\end{equation}
where $P_0$ is the measured mean probe power, and $m_p$ is the \emph{modulation index} of the microwave beatnote on the transmitted probe (see Eq. (\ref{eq:Pp})). This is true in coherent and incoherent optical detection schemes as shown in Appendix \ref{ap:D}. 

Suppose $\chi^{(1)} = \chi' -i\chi''$ is the dielectric susceptibility ``seen'' by the probe beam as it propagates a distance $L$ throughout the atomic ensemble. $\chi^{(1)}$ is dependent on the RF electric field magnitude at the transition frequency, and therefore will be modulated by the beatnote generated by the superposition of the LO and RF signal input.  The probe beam experiences a power attenuation $\exp\left(-2\kappa'' L\right)$, where $\kappa'' = (2\pi\nu_p/c)\sqrt{1+\chi'}\sqrt{1-i\chi''/(1+\chi')}$. In the absence of RF signal ${E_s}=0$, the probe suffers some non-zero attenuation which then becomes modulated by the presence of a small RF signal ${E_s} = \abs{E_s}e^{i\varphi}\neq 0$. This implies we can write $\kappa'' = \kappa''_0 + \alpha E_b(t)$, where $E_b(t)=\abs{E_s(t)}\cos\left(\omega_\mathrm{IF}t + \varphi(t)\right)$ and $\alpha$ some constant dependent on the transition dipole moment, EIT linewidth and chosen microwave LO amplitude. Hence, in the small signal regime, the transmitted probe power for a given input power $P_\mathrm{in}$ becomes

\begin{multline}
    {P_p(t)}{} = P_\mathrm{in}e^{-2\kappa_0'' L}e^{-2\alpha L E_b(t)}\\ \approx \underbrace{P_\mathrm{in} e^{-2\kappa_0'' L}}_{P_0}\left[1-2\alpha L E_b(t)\right]. \label{eq:Ppkappa}
\end{multline}
%
Here, $P_0$ is the mean probe power incident to the photodetector. For optically thin setups where $\kappa_0''L\ll 1$, $P_0$ is approximately independent from $L$. In that case, comparing Eq. (\ref{eq:Ppkappa}) with Eq. (\ref{eq:Pp}), we conclude via Eq. (\ref{eq:NEFtrend}) that $\mathrm{NEF}_0\propto \left(L\sqrt{P_0}\right)^{-1}$. For a Gaussian probe beam with fixed peak electric field magnitude determined by the optimal Rabi frequency, the optical power scales as $P_0 \propto w_0^2$, where $w_0$ is the beam waist. Therefore, in terms of the beam radius $w_0$ and interaction length $L$ 

\begin{equation}
    \mathrm{NEF}_0\propto \frac{1}{Lw_0}.\label{eq:NEFvolumetrend}
\end{equation}
Equation (\ref{eq:NEFvolumetrend}) agrees with the sensitivity dependence on interaction volume reported in \cite{meyerPhysRevA.104.043103}.

\section{NEF limits due to optical detection scheme}\label{ap:D}	


The power of the optical probe leaving the vapor cell $P_p(t)$ is modulated by the beatnote originating from the superposition of microwave LO at $\omega_\olo$ (zero-phase reference) and microwave signal at $\wsig=\omega_\olo + \omega_\mathrm{IF}$. If the electric field of the microwave signal has complex amplitude $E_s(t)=\abs{E_s(t)}e^{i\varphi(t)}$, then

\begin{equation}
    P_p(t) = P_0\left[ 1+ m_p  \underbrace{\abs{E_s(t)}\cos(\omega_\mathrm{IF} t + \varphi(t))}_{E_b(t)}\right].\label{eq:Pp}
\end{equation}
The slope $m_p$ is proportional to the dipole moment of the microwave transition and inversely proportional to the EIT linewidth. $P_0$ is the DC probe power incident towards the photodetector. Assuming a small-signal regime, i.e., $m_p\abs{E_s(t)}\ll 1$, and neglecting dispersion, the complex amplitude of the probe light also admits the linear approximation:

\begin{equation}
    A_p(t) = e^{i\varphi_p}\sqrt{P_0}\left[1 + \frac{m_p}{2}
    E_b(t)\right],\label{eq:Ap}
\end{equation}
where $\varphi_p$ is some time-independent phase.

\subsection{Direct (incoherent) optical detection}

Suppose the probe light at frequency $\nu_p$ is incident towards a photodetector of quantum efficiency $\eta$ and responsivity $R=q_e\eta/(h\nu_p)$, where $q_e$ is the electron charge. Using Eq. (\ref{eq:Pp}) the photocurrent $i_p(t)$ results

\begin{equation}
    i_p(t) = RP_0 + RP_0m_p{E_b(t)} + i_q(t) + i_n(t),
\end{equation}
where $\mathrm{var}\left[i_q(t)\right] = 2R^2\eta^{-1}h\nu_pP_0\Delta f$ is the photon shot noise \cite{loudon2000quantum,haus2000electromagnetic}, and $\mathrm{var}\left[i_n(t)\right] = R^2\mathrm{NEP}^2\Delta f$ the photodiode noise,  with $\Delta f$ the observation bandwidth, and $\mathrm{NEP}$ the photodiode's noise-equivalent power (in $\mathrm{W/\sqrt{Hz}}$). Both $i_q(t)$ and $i_n(t)$ are zero-mean processes with white spectrum.

One would estimate the beatnote electric field $E_b(t)$ by measuring the AC component of the photocurrent and dividing by $RP_0m_p$. However, that estimation has noise, whose standard deviation is $m_p^{-1}\sqrt{2\eta^{-1}h\nu_p\Delta f/P_0+\mathrm{NEP}^2\Delta f/P_0^2}$. After normalizing by $\sqrt{\Delta f}$, this quantity represents the noise equivalent field because it is the beatnote amplitude that is at the same level of the $1\sigma$ noise. Hence,

\begin{equation}
\mathrm{NEF} =\frac{1}{m_p} \sqrt{\frac{2h\nu_p}{\eta P_0}+\frac{\mathrm{NEP}^2}{P_0^2}}.\label{eq:NEFinc}
\end{equation}
%



\subsection{Homodyne optical detection}

After overlapping the probe with a strong optical local oscillator with field \footnote{Normalized to $\sqrt{\mathrm{W}}$ units such that $\abs{A_\mathrm{LO}}^2$ yields the total power of the LO.}  $A_\mathrm{LO}=\abs{A_\mathrm{LO}}e^{i\varphi_\mathrm{LO}}$ at the same frequency $\nu_p$, and measuring the result with a balanced photodiode, the differential photocurrent results (see Eq. (\ref{eq:Ap}))

\begin{equation}
\begin{split}
    i_p(t) &= 2R\Re{A_p(t)A_\mathrm{LO}^*} + i_q(t) + i_n(t) \\ 
        &= {i_{\mathrm{DC}}}+i_{q}(t)+i_{n}(t)\\&\quad +\underbrace{R\abs{A_{\mathrm{LO}}}\cos(\varphi_{d})\sqrt{P_{0}}m_{p}E_b(t)}_{i_{\mathrm{beat}(t)}}\label{eq:homodip}
\end{split}
\end{equation}
where $A_p(t)$ is given by Eq. (\ref{eq:Ap}), $\varphi_d=\varphi_p-\varphi_\mathrm{LO}$, $i_{\mathrm{DC}} = 2R\abs{A_{\mathrm{LO}}}\cos(\varphi_{d})\sqrt{P_{0}}$, $i_q(t)$ is a zero-mean Gaussian-distributed noise with variance $\mathrm{var}\left[i_q(t)\right]=  \frac{1}{2}h\nu_p\eta^{-1} R^2\abs{A_\mathrm{LO}}^2\Delta f$ \cite{shapiro,loudon2000quantum,haus2000electromagnetic}, and $i_n(t)$ has also zero mean and variance $\mathrm{var}\left[i_n(t)\right] = 2 R^2 \mathrm{NEP}^2 \Delta f$ where the factor of 2 arises from the subtraction of two statistically-independent photocurrent noises.  

One can estimate the beatnote electric field, by taking the AC part of $i_p(t)$ and dividing it by $R\abs{A_{\mathrm{LO}}}\cos(\varphi_{d})\sqrt{P_{0}}m_{p}$. The estimated signal is at the same level of the noise's standard deviation when

\begin{equation}
    \abs{E_{s}}=\frac{\sqrt{\frac{1}{2}h\nu_{p}\eta^{-1}R^{2}\abs{A_{\mathrm{LO}}}^{2}\Delta f+2R^{2}\mathrm{NEP}^{2}\Delta f}}{R\abs{A_{\mathrm{LO}}}\cos(\varphi_{d})\sqrt{P_{0}}m_{p}}.\label{eq:Eshomo}
\end{equation}
From this result we obtain the NEF after normalizing by $\sqrt{\Delta f}$, i.e., 

\begin{equation}
    \mathrm{NEF}=\frac{1}{\cos(\varphi_{d})m_{p}}\sqrt{\frac{h\nu_{p}}{2\eta P_{0}}+\frac{2\mathrm{NEP}^{2}}{P_{0}\abs{A_{\mathrm{LO}}}^{2}}}.\label{eq:NEFho}
\end{equation}

It can be seen that one advantage of homodyne detection is that the effect of thermally-induced current noise readout (NEP) can be arbitrarily reduced by increasing the optical LO power, leading to a quantum-limited detection process in a noisy photodiode. If the photodetectors' NEP is negligible for both homodyne and incoherent schemes, then homodyne detection is twice as sensitive. A homodyne detection scheme must guarantee $\varphi_d=0$ to minimize NEP.



\subsection{Heterodyne optical detection}

If the optical LO is detuned by some intermediate frequency $\nu_\mathrm{IFopt}$ (heterodyne optical detection), then an oscillating term $\exp(i2\pi \nu_\mathrm{IFopt}t)$ appears inside the real-value operator in Eq. (\ref{eq:homodip}). The effects of quantum noise and photodiode NEP are doubled with respect to homodyne detection because the observation bandwidth is not baseband but is instead centered at $\pm\nu_\mathrm{IFopt}$ in a dual-sided spectrum. Another way to interpret this is that the optical detection scheme is sensitive to both in-phase and quadrature components of noise, in contrast to homodyne detection. Therefore

\begin{equation}
 \mathrm{NEF}=\frac{1}{m_{p}}\sqrt{\frac{h\nu_{p}}{\eta P_{0}}+\frac{4\mathrm{NEP}^{2}}{P_{0}\abs{A_{\mathrm{LO}}}^{2}}}.   \label{eq:NEPhetero}
\end{equation}
When the photodiode's NEP is negligible, this implies a factor of $\sqrt{2}$ better than incoherent detection and $\sqrt{2}$ times worse than homodyne detection. However, in heterodyne detection no phase control is needed to maintain $\varphi_d=0$ because only the amplitude of the optical heterodyning beatnote is sufficient to retrieve amplitude and phase information from the RF signal.

\section{NEF limits from theoretical model}\label{ap:ModLim}
	
	We calculate NEF$_0$ from the model used to generate Figs. \ref{fig:sencont} and \ref{fig:compa10x} assuming photon-shot-noise of the probe beam is the only source of noise and a perfect photodetector. The simple direct optical detection scheme described in Sec. \ref{ap:D}.1. is assumed. The following calculations all assume an integration time of $\tau_\mathrm{int}=1\,\mathrm{s}$. 
	
	The probe power exiting the vapor cell and incident on a detector is
\begin{equation}
    P_p = P_\mathrm{in}T_r,
\end{equation}
	where $T_r$ is the transmission through the vapor found by the model. For a photon energy of $h\nu_p$ the photon-shot-noise is
\begin{equation}
	P_N = \sqrt{\expval{P_p} /(h\nu_p)} \cdot h\nu_p = \sqrt{\expval{P_p} h\nu_p},
\end{equation}
    where $\expval{P_p}=\expval{T_r}P_\mathrm{in}$ is the mean photodetected power. 
    Incident signals create a free-space interference beat-note with the local oscillator field, causing slowly varying changes in RF field amplitude and Rabi frequency $\Omega_\mathrm{RF}$. This produces a change in the transmitted probe power:
\begin{equation}
    \Delta P_p = P_\mathrm{in} \cdot \frac{\delta T_r}{\delta \Omega_\mathrm{RF}} \cdot \Delta \Omega_\mathrm{RF},
\end{equation}
    where $\delta T_r / \delta \Omega_\mathrm{RF}$ is the local linear slope of $T_r(\Omega_{RF})$ found from the model.
    
    The minimum detectable change in RF Rabi frequency, $\Delta \Omega_\mathrm{RF,min}$ is found by equating the change in transmitted power to the noise power:
\begin{equation}
    P_\mathrm{in} \cdot \frac{\delta T_r}{\delta \Omega_\mathrm{RF}} \cdot \Delta \Omega_\mathrm{RF,min} = \sqrt{\expval{T_r} P_\mathrm{in}h\nu_p},
\end{equation}
and solving for $\Delta \Omega_\mathrm{RF,min}$ 
\begin{equation}
    \Delta \Omega_\mathrm{RF,min} = 
    \sqrt{\expval{T_r}\frac{h\nu_p}{P_\mathrm{in}}} \cdot \frac{\delta\Omega_{RF}}{\delta T_r}. \label{eq:DOmegaRFmin}
\end{equation}
    Comparing Eq. (\ref{eq:DOmegaRFmin}) with Eq. (\ref{eq:NEFinc}) for an ideal photodetector, we conclude that $m_p={\delta T_r}/{\delta\Omega_{RF}}$ because $\Delta f = 1/(2\tau_\mathrm{int}) = 0.5\,\mathrm{Hz}$.  Using dipole matrix elements $\mu_d$ for the transitions $\ket{nS_{1/2,1/2}} \leftrightarrow \ket{nP_{3/2,3/2}}$ calculated by the ARC package \cite{SIBALIC2017319} the $\mathrm{NEF}_0$ is obtained as: 
\begin{equation}
    \mathrm{NEF}_0 = \sqrt{2}\frac{h}{\mu_d}\Delta \Omega_\mathrm{RF,min}.
\end{equation}

	\bibliography{biblio}    
\end{document}